\documentclass[journal]{vgtc}                

\ifpdf
  \pdfoutput=1\relax                   
  \usepackage{graphicx}                
  \DeclareGraphicsExtensions{.pdf,.png,.jpg,.jpeg} 
\else
  \usepackage{graphicx}                
  \DeclareGraphicsExtensions{.eps}     
\fi%

\graphicspath{{figs/}{pictures/}{images/}{./}} 

\def\BibTeX{{\rm B\kern-.05em{\sc i\kern-.025em b}\kern-.08emT\kern-.1667em\lower.7ex\hbox{E}\kern-.125emX}}

\newcommand{\eat}[1]{}

\usepackage{mathptmx}                  
\usepackage{xcolor}
\usepackage[most]{tcolorbox}
\usepackage{latexsym}
\usepackage{amsfonts}
\usepackage{amsmath}
\usepackage{amssymb}
\usepackage{colortbl}
\usepackage{epsfig}
\usepackage{xspace}
\usepackage{graphicx}
\usepackage{subfigure}
\usepackage{paralist}
\usepackage{enumerate}
\usepackage{hyperref}
\usepackage{cleveref}
\usepackage[color,matrix,arrow,all]{xy}
\usepackage{comment}
\usepackage{booktabs}
\usepackage{balance}
\usepackage{stmaryrd}
\usepackage{pifont}
\usepackage{hhline}
\usepackage{listings}
\usepackage{array}
\usepackage{float}
\usepackage[flushleft]{threeparttable}
\usepackage{enumitem}
\usepackage{wrapfig}

\newcolumntype{L}[1]{>{\raggedright\let\newline\\\arraybackslash\hspace{0pt}}m{#1}}
\newcolumntype{C}[1]{>{\centering\let\newline\\\arraybackslash\hspace{0pt}}m{#1}}
\newcolumntype{R}[1]{>{\raggedleft\let\newline\\\arraybackslash\hspace{0pt}}m{#1}}

\usepackage{tikz}

\usepackage{epsfig}
\usepackage{multirow}
\usepackage{url}

\usepackage[export]{adjustbox}
\usepackage[noend]{algorithmic}
\usepackage{algorithm}

\newcommand{\red}[1]{#1}

\newcommand*\circled[1]{\tikz[baseline=(char.base)]{
\node[shape=circle,fill=red,inner sep=1pt] (char) {\textcolor{white}{\small #1}};}}
\newcommand*\bluecircled[1]{\tikz[baseline=(char.base)]{
\node[shape=circle,fill=blue,inner sep=1pt] (char) {\textcolor{white}{\small #1}};}}

\makeatletter
    \newcommand\figcaption{\def\@captype{figure}\caption}
    \newcommand\tabcaption{\def\@captype{table}\caption}
\makeatother

\definecolor{yellow}{HTML}{FFFF00}
\definecolor{darkgreen}{rgb}{0.15,0.55,0.15}
\definecolor{darkblue}{rgb}{0.1,0.1,0.5}
\definecolor{blue}{RGB}{68,114,196}
\definecolor{darkgreen}{rgb}{0.15,0.55,0.15}
\definecolor{mred}{rgb}{.80,.12,.30}
\definecolor{grey}{rgb}{0.5,0.5,0.5}
\definecolor{Purple}{rgb}{.75,0,.85}
\definecolor{light-gray}{gray}{0.95}
\definecolor{mid-gray}{gray}{0.85}
\definecolor{darkred}{rgb}{0.7,0.25,0.25}

\newcommand{\blue}[1]{\textcolor{blue}{#1}}

\newtheorem{myExample}{Example}

\newtheorem{example}[myExample]{Example}

\newcommand{\stitle}[1]{\vspace{1ex}\noindent{\bf #1}}

\setitemize{noitemsep,topsep=0pt,parsep=0pt,partopsep=0pt,leftmargin=*}

\newtcbox{\mywboxtext}{on line,
colback=yellow,frame hidden,colframe=white,size=fbox,boxrule=0pt,fontupper=\color{red}}
\newcommand{\hl}[1]{\textrm{#1}}

\def\ojoin{\setbox0=\hbox{$\Join$}%
\rule[-0.05ex]{.27em}{.4pt}\llap{\rule[1.3ex]{.27em}{.4pt}}}
\def\leftouterjoin{\mathbin{\ojoin\mkern-5.8mu\Join}}

\def\fullouterjoin{\mathbin{\ojoin\mkern-5.8mu\Join\mkern-5.8mu\ojoin}}
\onlineid{0}
\vgtccategory{Research}
\vgtcpapertype{Representations \& Interaction}
\title{View Composition Algebra for Ad Hoc Comparison}

\newcommand{\sys}[0]{\texttt{VCA}\xspace}

\author{Eugene Wu}
\authorfooter{
\item Eugene Wu is with Columbia University.  Email: ewu@cs.columbia.edu.
}

\shortauthortitle{Wu, E: View Composition Algebra for Ad hoc Comparison}


 \teaser{
   \centering
   \includegraphics[width=.95\linewidth]{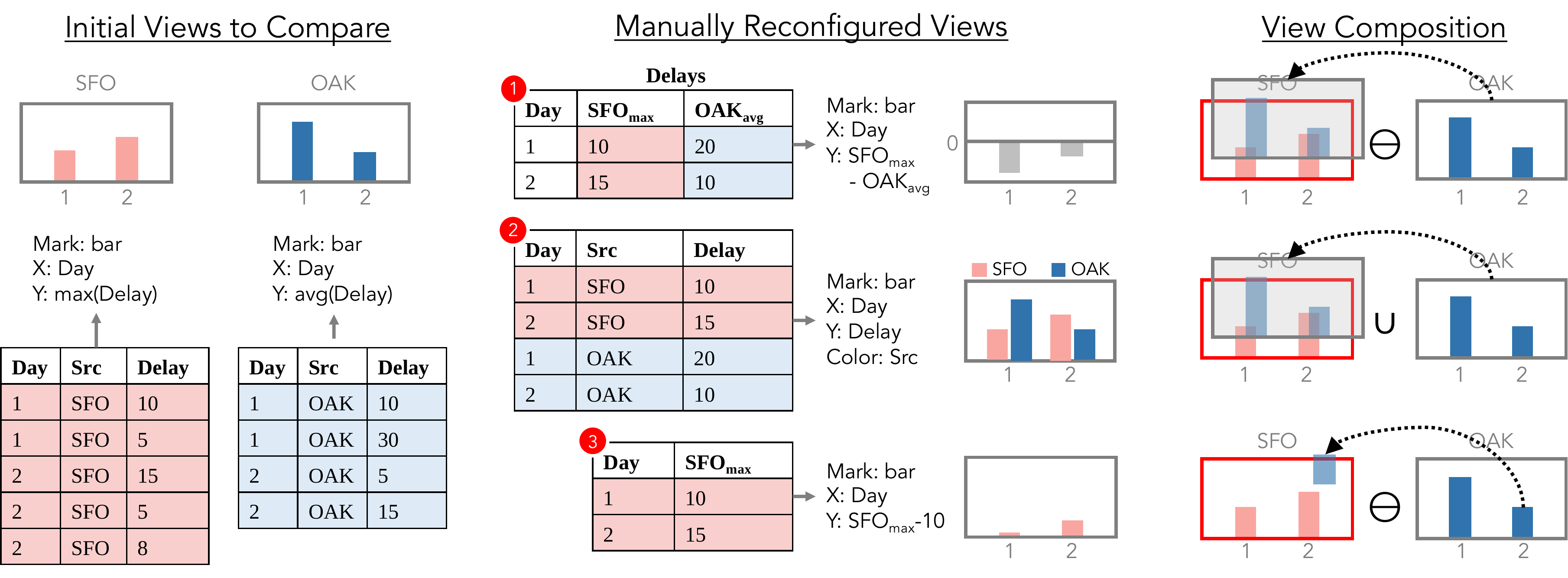}
   \vspace{-1em}
   \caption{
     Comparing two flight delay visualizations.
   (Left) Data and transformations to render \texttt{SFO}'s maximum daily delay and \texttt{OAK}'s average daily delay.  (Middle) Comparing data between the two charts requires data transformations that are hard or not possible to express in visualization systems and languages: 1) compare SFO and OAK statistics for each day, 2) juxtaposing statistics of the same days, and 3) compare OAK's second day with both of SFO's days.  (Right) Visual Composition Algebra (\sys) easily expresses these comparisons as the difference ($\ominus$) and union ($\cup$) between the two charts, and the difference between the \texttt{SFO} chart and a mark, respectively.  \sys  is a new formalism that defines 7 operators to make ad hoc comparisonsbetween values, marks, legends, and charts.    }
 	\label{f:teaser}
 }

\vgtcinsertpkg

\abstract{
Comparison is a core task in visual analysis.  Although there are numerous guidelines to help users design effective visualizations to aid known comparison tasks, there are few techniques available when users want to make ad hoc comparisons between marks, trends, or charts during data exploration and visual analysis.  For instance, to compare voting count maps from different years, two stock trends in a line chart, or a scatterplot of country GDPs with a textual summary of the average GDP.  Ideally, users can directly select the comparison targets and compare them, however what elements of a visualization should be candidate targets, which combinations of targets are safe to compare, and what comparison operations make sense?  This paper proposes a conceptual model that lets users compose combinations of values, marks, legend elements, and charts using a set of composition operators that summarize, compute differences, merge, and model their operands.  We further define a View Composition Algebra (\sys) that is compatible with datacube-based visualizations, derive an interaction design based on this algebra that supports ad hoc visual comparisons, and illustrate its utility through several use cases.
}

\begin{document}
\maketitle

\section{Introduction}\label{s:intro}

Comparison is ubiquitous in visual data analysis, as a low-level perceptual task~\cite{cleveland1984graphical,Talbot2014FourEO,Heer2012InteractiveDF} as well as a high-level analysis task~\cite{Munzner2014VisualizationAA,gleichercompare}.  When the comparsion targets and tasks are known apriori, there are numerous design guidelines~\cite{Javed2012ExploringTD,gleichercompare} to help designers choose effective visual encodings and visualization designs.  But what if a visualization end-user wants to perform ad hoc comparisons during visual analysis?

\begin{figure}
\centering
\includegraphics[width=\columnwidth]{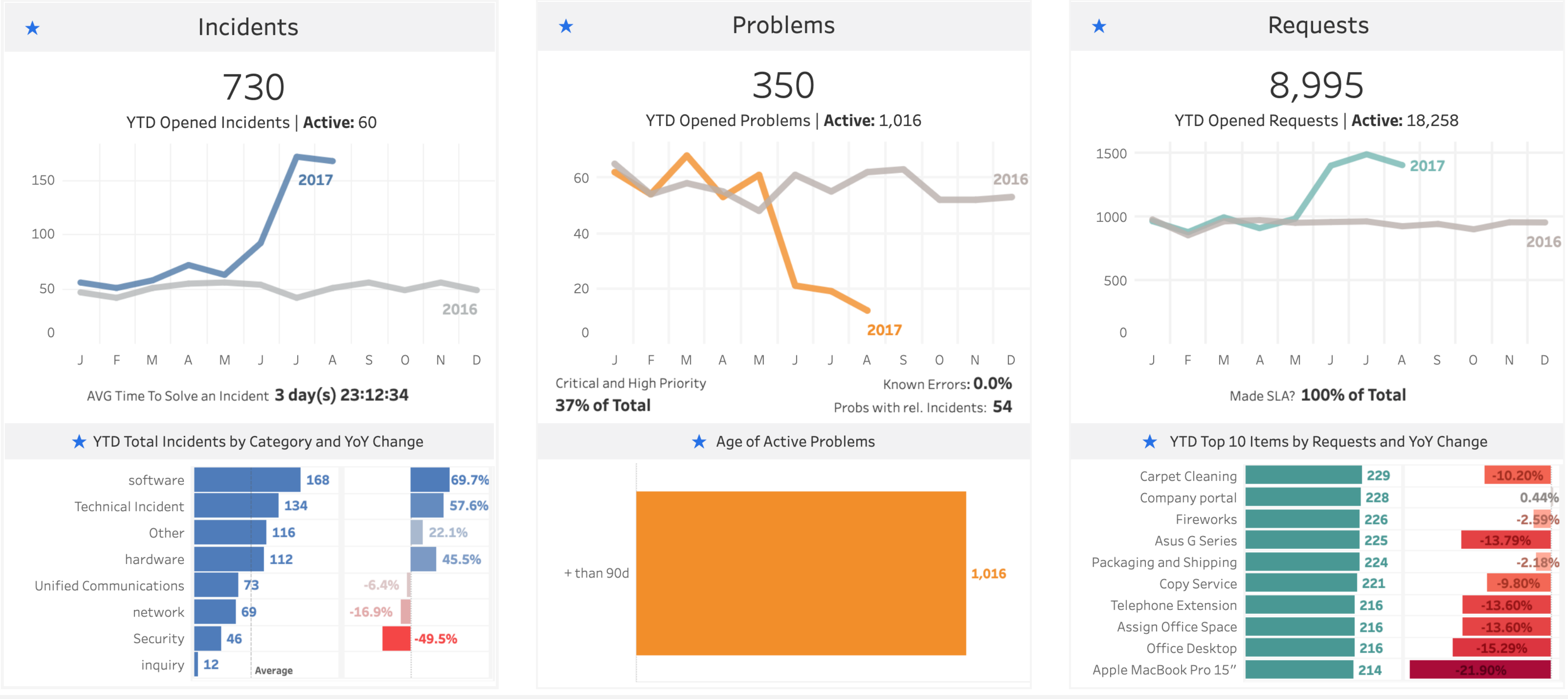}
\vspace{-.2in}
\caption{Tableau Online's pre-built ServiceNow  Executive Dashboard.}
\label{f:tableau}
\end{figure}

\Cref{f:tableau} is Tableau's example ServiceNow executive dashboard and shows the number of incidents, problems, and requests per year (top) and their breakdowns (bottom).  But the user may be interested in comparisons the dashboard was not design for---what if the user wants to know the difference between the 2017 and 2016 incident trends, or compare them with the year-to-date hardware incidents, or compare any other combination of values, marks, trends, or charts?  Although the user can see and point to the comparison targets, existing visualizations do not perform and visualize the comparison.  In practice, users visually estimate comparisons (across encodings or even charts), and risk making biased or inaccurate estimates~\cite{Talbot2014FourEO,cleveland1984graphical,Zeng2021AreWT}.

It would be useful to develop a set of comparison-specific interaction techniques that visualization developers and visual analysis systems can readily incorporate.  When designing a new class of interactions, it is helpful to build the interaction techniques on top of a formal language. This decouples the interaction design from the formal semantics, so that a single language implementation can support a multitude of interaction designs.  Examples of such languages include table algebras~\cite{wilkinson2006grammar,stolte03thesis} for facetted layouts, graphical grammars~\cite{wickham2016ggplot2,wilkinson2006grammar} for statistical charts; and interaction grammars~\cite{Satyanarayan2017VegaLiteAG} for manipulation and coordination interactions.     

Although no formalism has been developed for comparison, it is sensible that any formalism should let users interactively compare any visual representation of data (e.g., charts, marks) as targets, as long as they are comparable.  This naturally leads to five desirable criteria regarding the nature of the input targets (C1,2), the output comparison results (C3), and what comparable means (C4,5).

\begin{itemize}[leftmargin=*,itemsep=0in]
  \item {\it (C1) Flexible Targets:} comparison targets should be any visualization component derived from data, such as constant values, marks, legend elements, and charts.
  \item {\it (C2) Design Independence:} whether targets are comparable should be independent of their visual encoding, design, and spatial placement.  For instance, the per-category incidents in the bar chart should be comparable with the monthly incidents line chart despite their visual differences.
  \item {\it (C3) Design Diversity} the comparison result should support common visual design strategies for comparison~\cite{gleichercompare}, namely superposition, juxtaposition, and explicit encoding of the differences.
  \item {\it (C4) Safety:} of the possible combinations of targets, only a subset are unambiguous and semantically meaningful to compose.  For instance, \texttt{profits} and  \texttt{height} are quantitative attributes but should not be composed.  The formalism should identify comparisons between targets that are potentially unsafe, and either prevent it or warn the user.  
  \item {\it (C5) Expressiveness:} targets that are semantically compatible should be allowed.
\end{itemize}

\noindent C4 and C5 they pose an interesting question: {\it When are visualization data comparable?} Traditionally, the visualization designer manually determines that data is safe to compare, and then chooses the appropriate data transformation and design strategies.  However, ad hoc comparisons require that we formalize safety to check it automatically.  The challenge is that safety is trivial to satisfy by rejecting all comparisons.  However, widening expressiveness is nontrivial because safety depends on the view definition, dataset's semantics, and user's domain knowledge.   Thus, a comparison language must strive to maximize expressiveness while ensuring safety.

\smallskip
\noindent{\bf Contributions: } This paper makes three core contributions: a conceptual model for comparison operations, a concrete algebra called View Composition Algebra (\sys\footnote{\sys is also short for Visual Comparison Algebra.}) that  satisfies C1-5 and is designed for multi-dimensional cube-based analytics~\cite{Gray2004DataCA}, and an interaction design based on the algebra.  

The conceptual model shows that comparison differs from typical interaction techniques in that it must understand the semantics of the view's entire data processing pipeline.  The model also introduces the notions of safety and expressiveness for comparison interactions.

\sys introduces a closed set of composition operators over any visual representation of data, which we call {\it Views}\footnote{{\it View} in data management refers to a query result; here it refers to the visual representation of a query result.}, including labels, marks, and charts.  The algebra defines three types of view composition operators that computes statistics, merges marks, or fits statistical models to the data from the input views.  We describe how to compile statements into SQL, and show that \sys is compatible with visualization formalisms such as VizQL~\cite{Stolte2000PolarisAS}.   We further highlight subtle challenges that arise when comparing views that are compatible but ``mis-aligned'' (e.g., comparing prices on even vs odd days) and introduce a novel {\it lift} operator to facilitate such comparisons.

The interaction design lets users select values, marks, legend elements, or charts as comparison targets, and choose composition operators to apply to the targets.  We illustrate how comparison interactions can express computations otherwise not possible in existing visualization systems, enable comparisons across visual encodings, can be composed to perform multi-step comparisons, and are compatible with existing visualization interactions such as dynamic filtering.

The next section characterizes comparison interactions within the context of existing interaction taxonomies, and provides background on comparison and language designs.   The subsequent sections introduce the conceptual model, algebra, interaction design, and case study.  We end by discussing the limitations and potential research directions to expand on the ideas in this paper.

If there is one takeaway message, it is that ad hoc visual comparison depends on analyzing {\it data-level} transformations and analyses {\it as well as} visual design principles. \sys is a starting point in the interaction design space for view composition and comparison.   The text uses screenshots when possible to describe the interaction design, and simplified diagrams when they are more legible.  Demos and code can be found at \url{http://viewcompositionalgebra.github.io}.

\section{Background and Related Work}

This section first places comparison interactions in the context of existing interactions in Yi et al.'s~\cite{Yi2007TowardAD} taxonomy, and then contrasts with existing work on comparison.

\begin{figure}
\centering
\includegraphics[width=.85\columnwidth]{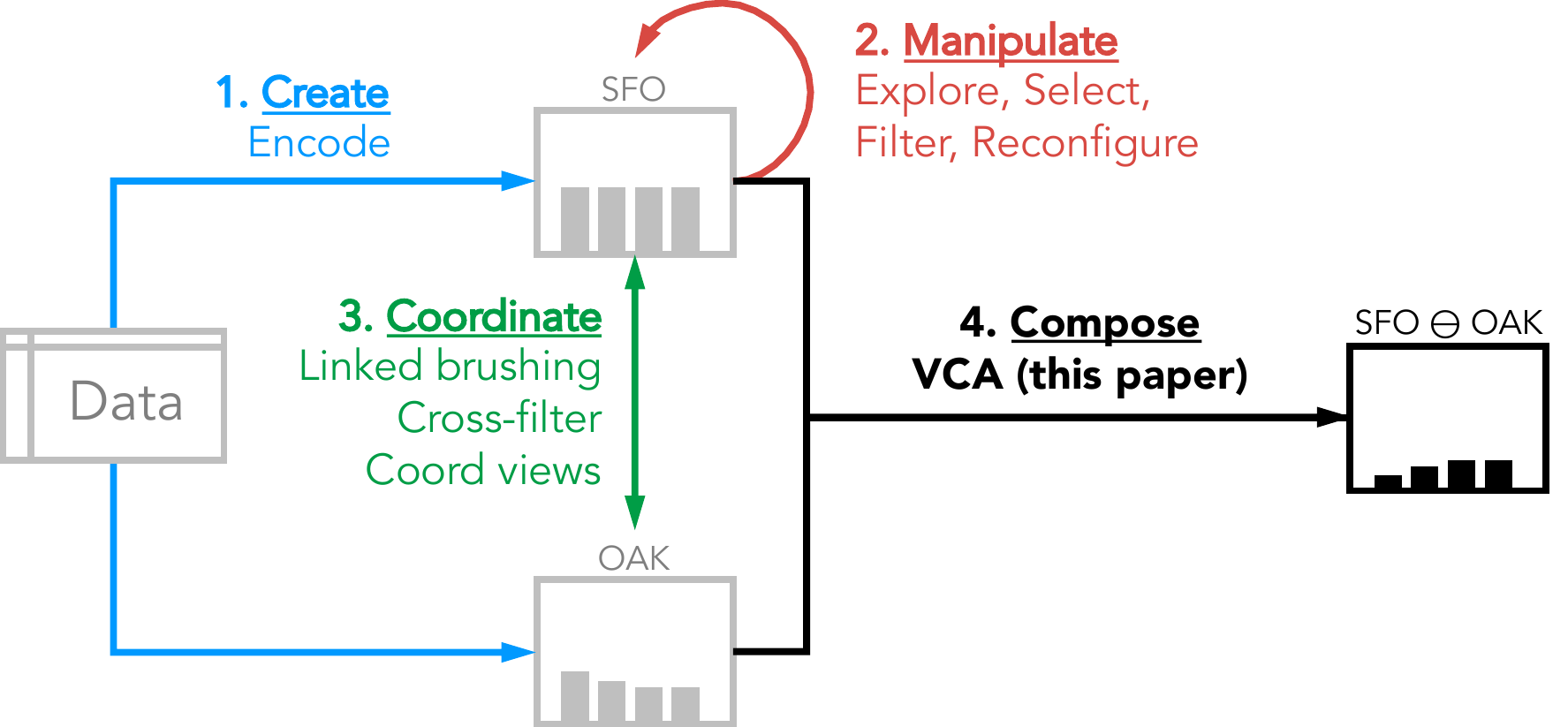}
\caption{Categorization of interactions in visual analysis: (1) View creation, (2) Single-view manipulation, (3) Multi-view coordination, and (4) View composition.  Existing formalisms support (1-3); this paper proposes a formalism for (4) to support comparison.}
\label{f:layofland}
\end{figure}

\subsection{Interaction Taxonomies}

Historically, comparison has been treated as a high level goal, rather than an interaction type.  For instance, Yi et al.~\cite{Yi2007TowardAD} omit {\it Compare} as a category because it can be achieved via e.g., filtering, arrangement, reconfiguration, or encoding.  Similarly, Munzner's multi-level task taxonomy~\cite{Munzner2014VisualizationAA} places comparison under {\it Why} as a low-level analysis query, rather than as a mechanism under {\it How}.  

Thus, a helpful way to contextualize comparison as an interaction is to model interactions as functions over data and their visual representations ({\it Views}).  We can then organize Yi et al.'s~\cite{Yi2007TowardAD} taxonomy of interactions (in {\it italics} below) based on the number of views the interaction affects and the nature of the interaction (\Cref{f:layofland}).  

\begin{itemize}[leftmargin=*,itemsep=0in]
  \item {\it Create:} uses a dataset to create new views based on their visual {\it encoding}~\cite{slingsby2009configuring,Stolte2000PolarisAS,wickham2016ggplot2,Satyanarayan2017VegaLiteAG,Wu2017CombiningDA}. 

  \item {\it Manipulate:} involves interactions within a single view, such as {\it select}~\cite{Satyanarayan2017VegaLiteAG,Tukey2017ExploratoryDA}, {\it explore}~\cite{Shneiderman1996TheEH}, or {\it filter}~\cite{Shneiderman1994DynamicQF}.  It also includes interactions that {\it reconfigure} the view's visual encodings or data transformations.  

  \item {\it Coordinate: } involves coordination between multiple views, where an interaction (typically selection) within one view will highlight~\cite{Tukey2017ExploratoryDA}, filter, or otherwise update data~\cite{Weaver2004BuildingHV,north2000snap} in linked views.  The relationships are typically pair-wise, so that each linked view is updated independently.

      \item {\it Compose: } takes multiple views as input to derive one or more output views.  The dominant interaction type {\it rearranges} the spatial layout of multiple views or layers views atop each other~\cite{Javed2012ExploringTD,Baldonado2000GuidelinesFU,Roberts2007StateOT,Chen2020CompositionAC,Satyanarayan2017VegaLiteAG}.  Rearrangement is primarily a design-level operation.
    
  \end{itemize}

\noindent Comparison is a composition operation, however it differs from rearrangement in that it analyzes the data- and design-level specifications of the input views to derive the output views.  Existing interactions---such as re-encoding data in different views to share an axis~\cite{Qu2018KeepingMV}, or rearranging views to superimpose them atop each other---help make comparison {\it less difficult}, however none directly perform comparison.  

Some visualization systems such as Tableau support manual data transformations, and can technically express comparison tasks.  However, even simple comparisons require complex steps that users find difficult and confusing~\cite{tableauhelp1,tableauhelp2,tableauhelp3,tableauhelp4}.
To illustrate these challenges, the following example shows two simple bar charts of flight delays out of SFO and OAK airpoirts, and the very different data tables that a user would need to manually construct in order to render simple comparisons between the two charts.  It also shows how \sys would let users express these comparisons interactively.

\begin{example}
  \Cref{f:teaser}(left) shows the input data, visual encodings, and charts of the maximum and average flight delays by day from \texttt{SFO} and \texttt{OAK}, respectively.   The middle column shows the transformed data and encodings needed to express three similar comparisons: 1) subtract \texttt{OAK} from \texttt{SFO} statistics by day, 2) show \texttt{SFO} and \texttt{OAK} statistics side by side, and 3) compare \texttt{SFO}'s statistics with \texttt{OAK}'s average delay on day 2.  However, the transformations and resulting tables in each case are quite different.  The first aligns \texttt{SFO} and \texttt{OAK} statistics by day and creates separate attributes for each; the second concatenates the \texttt{SFO} and \texttt{OAK} statistics; and the third uses \texttt{SFO}'s statistics as the input table, and updates the formula in the visual mapping.
\end{example}

\subsection{Graphical Perception}

Comparison is a core judgement task in graphical perception.  Participants in Cleveland and McGill's original studies~\cite{cleveland1984graphical}, as well as numerous later studies~\cite{Talbot2014FourEO,Heer2010CrowdsourcingGP,Zacks1998ReadingBG,Zeng2021AreWT} compared two graphically encoded values, and used judgement accuracy to rank different encodings.  These studies find that distance, encoding, distractors, and even the values themselves affect judgement accuracy.  Interestingly, interaction techniques ``{\it that bring distant bars closer together, such as sorting, drawing reference lines, or windowing}''~\cite{Talbot2014FourEO} have been proposed to minimize potential judgement errors.  While these may reduce errors, interactions that directly compute the comparison can help avoid many of these errors altogether.

\subsection{Design Strategies for Comparison}

To the best of our knowledge, Tominski et al.~\cite{Tominski2012InteractionSF} were the only to propose a general set of interactions to compare visualizations based on how users juxtapose (place them side by side) and superpose (placed atop each other, or flipping between pages) graphs printed on sheets of paper.   They implement digital versions of these interactions.  Similar to this work, users can select subviews to compare.  However, they treat views as generic ``pieces of graphical information'' and explicitly ``abstract from the specific details of the data to be compared'' because ``developing general support for visual comparison [is] a challenging endeavor.''  This work formalizes such general support by explicitly considering the data details.  

Javed and Elmqvist~\cite{Javed2012ExploringTD} propose a design space for composite visualizations, describe four visual composition designs (juxtapose, superpose with and without shared axes, and nested views), and highlight their benefits and drawbacks.  
Gleicher et al.~\cite{Gleicher2011VisualCF} survey over 110 visualization comparison papers, and organize them under a taxonomy of juxtaposition, superposition, and explicit encoding strategies. Their later work~\cite{gleichercompare} characterizes comparison by the target elements being compared, the actions taken to compare the elements, and one of three design strategies from their taxonomy.  Both Javed et al. and Gleicher et al. focus on design guidelines when comparison tasks are known, but do not address the data transformation and safety aspects.  We argue that the latter is crucial for ad hoc comparison, and design \sys to support all three of Gleicher et al.'s design strategies.

Qu and Hullman~\cite{Qu2018KeepingMV} formalize visual consistency properties when the same data attributes are visually encoded in multiple views, and study when designers violate these guidelines.  Consistent scales can improve visual comparison, even when marks are separated, but are subject to the same issues in graphical perception above.  Similarly, Kinneman and Scheidegger's AlgebraicVis~\cite{Kindlmann2014AnAP} models visualizations as a sequence of data and visual transformations, and defines invariants that should hold under changes to the data or visual respresentations.  Such invariants can also be interpreted as a consistency measure.

\subsection{Composition in Visualization Grammars}

Graphical grammars, such as ggplot2~\cite{wickham2016ggplot2}, Vega-lite~\cite{Satyanarayan2017VegaLiteAG}, and VizQL~\cite{Stolte2000PolarisAS} model visualizations as mappings from data attributes to spatial layout characteristics (e.g., facetting) as well as visual attributes.  Although users can explicitly specify data transformations, systems built on these languages also perform implicit data transformations (e.g., default aggregation functions, ``nice'' binning) that are not reflected in the specification, and can make analysis challenging.  These grammars support a form of  composition that spatially organizes views into a multi-view visualization.  For instance, Vega-lite's composition operators are used to layer, facet, concatenate, and repeat views.  However, these alone are not sufficient for many comparison tasks that require data-level analysis and transformations.  In addition, layering is a form of composition, where safety considerations also arise (\Cref{s:discussion}).

\subsection{Visual Analysis Systems}

Academic~\cite{Weaver2004BuildingHV,livny1997devise,north2000snap,derthick1997interactive,aiken1996tioga,Satyanarayan2017VegaLiteAG,wickham2016ggplot2,wilkinson2006grammar} and commercial~\cite{tableau,spotfire,powerbi,sigmacomputing} visual analysis systems are largely designed as client-server applications, where the client translates user interactions into SQL queries and renders the query results as visualizations in the interface.  Since manually implementing this translation layer is possible but tedious, systems use visualization algebras and grammars~\cite{Stolte2000PolarisAS,wilkinson2006grammar,wickham2016ggplot2} support a rich set of interaction designs and easily translate into database queries.  \sys is a meta-language on top of these queries to compare data in these visualizations.

Coordinated Multi-view Visualizations~\cite{Roberts2007StateOT} (CMVs) have also been used to aid comparison by rendering different perspectives of the same data, such as overview-detail, focus-context, and primary-secondary.  This helps users switch from performing mental to visual comparison tasks~\cite{Baldonado2000GuidelinesFU}.  Of note are difference views that explicitly show differences between two views.  However, they are difficult to achieve in practice, and  primarily focused on textual data~\cite{Suvanaphen2004TextualDV,Seeling2004AnalysingAO}.
In a sense, creating new views through comparison interactions is a way to dynamically create a multi-view visualization.

\section{Conceptual Model and Algebra Overview}
\label{s:overview}

This section provides the conceptual model behind the View Composition Algebra (\sys) that supports the five desired criteria introduced in \Cref{s:intro}, its relationship with data integration, and its safety properties.  The next section describes a concrete instantiation of this algebra for multi-dimensional cube-based visualizations.

\subsection{Overview}\label{ss:vcaoverview}
We first describe our model of a view (operand), and then describe the main operators in \sys.  The key idea is to separate data transformation from rendering-specific operations, so that each can be analyzed and changed independently.  The focus on the data-level helps support flexible targets (C1) independently of their visual presentation (C2).   Note that we do not consider layout-oriented composition (e.g., rearranging views side by side) since it is well-studied~\cite{Chen2020CompositionAC} and widely supported.

\begin{table}
  \small
  \begin{minipage}[t]{.34\linewidth}
    \centering
    \parbox[t]{\linewidth}{
      \begin{tabular}{ccr}
        \multicolumn{3}{c}{\circled{1} \textbf{Chart}}\\
        Date&Src&\blue{Delay}\\
        \hline
        1&SFO&\blue{10}\\
        2&SFO&\blue{15}\\
        3&SFO&\blue{20}\\
        1&OAK&\blue{15}\\
        2&OAK&\blue{10}\\
        3&OAK&\blue{5}\\
      \end{tabular}
    }
  \end{minipage}
  \hfill
  \begin{minipage}{.63\linewidth}
    \begin{minipage}[t]{\linewidth}
    \centering
    \parbox[t]{.3\textwidth}{
      \begin{tabular}{c}
        \multicolumn{1}{c}{\circled{2} \textbf{Value}}\\
        \blue{Delay}\\\hline
        \blue{20}
      \end{tabular}
    }
    \hfill
    \parbox[t]{.68\textwidth}{
      \begin{tabular}{ccr}
        \multicolumn{3}{c}{\circled{3} \textbf{Mark(s)}}\\
        Date&Src&\blue{Delay}\\
        \hline
        1&SFO&\blue{10}
      \end{tabular}
    }
    \end{minipage}
    \begin{minipage}{\linewidth}
      \parbox[t]{.52\textwidth}{
        \begin{tabular}{ccr}
          \multicolumn{3}{c}{\circled{4} \textbf{Group (b=2)}}\\
          Date&Src&\blue{Delay}\\
          \hline
          1&OAK&\blue{15}\\
          2&OAK&\blue{10}\\
          3&OAK&\blue{5}
        \end{tabular}
      }
      \hfill
      \parbox[t]{.42\textwidth}{
        \begin{tabular}{ccc}
          \multicolumn{3}{c}{\circled{5} \textbf{Model}}\\
          \multicolumn{3}{c}{$f(Date,Src)\to \blue{Delay}$}\\
          &&\\
          &&
        \end{tabular}
      }
    \end{minipage}
  \end{minipage}
  \vspace{1em}
  \caption{Tabular representation of charts, constant values, marks, groups, and predictive models.
  Dimensions in black, measures in \blue{blue}}
  \label{t:viewtypes}
\end{table}
\begin{figure}[th]
  \centering
  \vspace{-1em}
  \includegraphics[width=.4\columnwidth]{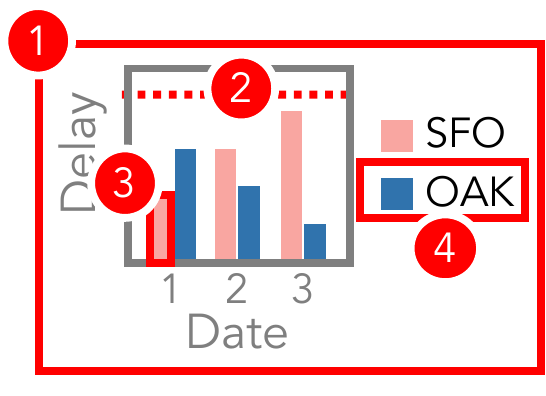}
  \vspace{-1em}
  \caption{Bar chart and its component views: constant value, mark, and grouping attribute in the legend.  Chart and components correspond to tables in \Cref{t:viewtypes}. }
  \label{f:viewtypes}
\end{figure}

\stitle{Operands/Views:} 
A {\it view} traditionally refers to an entire chart.  In contrast, \sys defines view such that any part of the visualization that conforms to this definition is a {\it view}. 

We model a view as the output of $R(T)$, where $T$ is a table (a raw table or a SQL query result) and $R$ is a visual encoding specification from a subset of $T$'s data attributes to visual attributes.  For example, \Cref{f:viewtypes} illustrates a bar chart that renders table \circled{1} in \Cref{t:viewtypes}, where $R$ is encodes \texttt{Date} to the x-axis, \texttt{src} to color, and the measure \texttt{Delay} to the y-axis.   We call $T$'s attributes its {\it schema}.   Similar to VizQL~\cite{Stolte2000PolarisAS}, the attributes in $T$ are either dimensions (used for filtering, grouping, and database joins) or measures (used to compute statistics); for presentation, we assume that each table contains a single measure.

In practice, $T = Q(D)$ is the output of data transformations $Q$ applied to an underlying database table $D$.  Thus, a view $V=R(Q(T))$ is the result of a query followed by visual encoding.  We assume that $Q$ encapsulates {\it all} computations and transformations needed to produce $T$, and the renderer simply maps data rows in $T$ to marks (or other objects) in the view.

We use \Cref{f:viewtypes} and \Cref{t:viewtypes} to illustrate the above concepts.  The chart is annotated with the different types of views that can be expressed as the tables listed in \Cref{t:viewtypes}.  
For instance, the chart's input table is depicted as \circled{1}. A constant value is a single-row table with a single attribute and rendered as the dotted line \circled{2}, and the highlighted bar in Date 1 is a row from the input table \circled{3}. A grouping attribute rendered in the legend represents a subset of rows that match a predicate: \texttt{OAK} corresponds to the rows that match \texttt{src=OAK} (\circled{4}).  Finally, predictive models use a subset of the dimensions  to predict a measure.  For instance, \circled{5} uses \texttt{Date} and \texttt{src} to predict the expected \texttt{Delay}.  In essence, $T$ is an instance of a table sampled from the model.  

In general, we require the view to maintain mappings between rendered objects and their underlying data rows---which objects have a mapping, and how the mappings are maintained is implementation-specific.  For instance, a line chart typically renders a linear interpolation between the input rows.  The view may implement functionalities that let the user e.g.,  select an individual point along a line to retrieve its input row, the entire line to retrieve all of its input rows, or a segment to retrieve that subset of rows. 

\stitle{Operators:} 
Composition uses views to derive new views.  As such, we define different unary, binary, and n-ary composition operators and two decomposition operators.  For instance, a binary operator $O(V_1,V_2)\to V^*$ derives a new view $V^*=R^*(Q^*(D))$ with a new visual encoding $R^*$ and query $Q^*$.  The new visual encodings are based on Gleicher's taxonomy of comparison design strategies~\cite{gleichercompare} (satisfying the diversity criteria (C3), and we use operator and input view characteristics to choose the strategy. 

The binary operators {\it statistical composition} and {\it union composition} are respectively used to derive new statistics and superimpose/juxtapose the operands.  The n-ary operators generalize the binary operators to compute aggregate statistics over, and superimpose/juxtapose, a set of views.  The unary {\it lift} operator transforms a view into a predictive model; this is useful when the views are technically safe to compose, but do not share the same attribute values.  For example, composing two views that render prices on even days and odd days is not meaningful  because each odd day does not have a counterpart to comparable against in the even days.   Lifting the even days view will interpolate its prices to odd days so that the composition produces a meaningful result.  Finally, the unary extract and explode operators are for view {\it decomposition}, where components of a view are extracted and turned into new standalone views.

\subsection{Composition Safety and Expressiveness}
\label{ss:safety}

Before defining {\it how} views can be composed, we must understand {\it when} data from multiple views are safe to compose.  We now draw a connection between data integration and composition safety, and use the observation to define an algebra's safety and expressiveness properties.  We also describe \sys' heuristics.

\subsubsection{Connection with Data Integration}

Data integration~\cite{doan2012principles,batini1986comparative,Naumann2018SchemaM,batini1986comparative,christen2012data} is a research field that studies the problem of combining two (or more) tables into a single ``integrated'' table, and deals with two primary problems. {\it Schema matching} maps attributes in one schema to semantically equivalent attributes in the second schema.  For instance, which attributes in the schemas \texttt{(day, price)} and \texttt{(date, cost)} should match?    {\it Entity matching} maps  rows (i.e., entities) in one table to the ``comparable'' rows (i.e., the same entities) in the second table.

The same two problems manifest in view composition.  Determining whether the views are composable is an instance of schema matching.  For instance, a chart of \texttt{profits} vs \texttt{losses} should clearly not be composed with a chart of \texttt{names} vs \texttt{age}. However, it would also be ambiguous to compose with a chart of \texttt{profits} vs \texttt{profits} (of different years) since it is unclear which \texttt{profits} attribute in the latter chart should be matched with \texttt{profits} in the first chart.  

Similarly, determining which marks in the input views should be compared is an instance of entity matching because we are asking which underlying data records (that each mark renders) should be matched and compared.    For example, suppose we wish to compose the chart in \Cref{f:viewtypes}, whose dates span $1-3$, with a different chart whose dates span $5-8$.  Although their schemas are identical, none of the marks in the first view can be compared with any marks in the second view because their dates do not overlap.

In short, the schema matching problem determines whether or not composition should be allowed (safety), whereas the entity matching helps us assess whether the composed view return a meaningful result.  In SQL terminology, the former determines if a join is possible, and the latter determines if the join would return any results.  This correspondence informs our definition and data-oriented approach to safety.   

Finally, both schema and entity matching are open problems with considerable and active interest in both  academia and industry.  This suggests that ad-hoc comparison is also difficult in general.  This motivates the heuristics that we use below, but also suggests that we can borrow techniques from these fields to improve upon our heuristics.

\subsubsection{Safety}

Safety is necessary to avoid ambiguous or misleading compositions.  For instance, a chart of \texttt{delay} by \texttt{date} should not be composed with a chart of \texttt{profits} by \texttt{market}.   \sys enforces safety by only allowing compositions for which there is a unique schema matching between the views' schemas (C4).  Since this is in general an open problem, our implementation uses a simple heuristic: two views are incompatible if their schemas cannot be unambiguously matched or if their measures are incompatible.   This heuristic could be replaced with learning- or rule-based approaches from the schema matching literature.  

For simplicity, two attributes are compatible if they are the same\footnote{Alternatives include classification models that predict if two attributes are compatible, or a rules that specify that e.g., \texttt{profit} and \texttt{cost} are compatible.}.  
Two tables match if, for each attribute in each table, there is a unique attribute in the other table that it is compatible with (they have the same set of dimensions).  For binary operators, this requirement only needs to hold for one table (dimensions have a superset-subset relationship, see (\Cref{s:algebra})).  Using this heuristic, the views in the above example do not match because their dimensions (\texttt{date}, \texttt{market}) do not overlap.   

Two measure attributes are compatible if they are derived from the same input attribute and use compatible statistical functions (if any) as determined by our attribute-sensitive type checking rules.  A function $f(a)$ can have output type $a$, a custom type $T$, or a parameterized custom type ${T}{<}{a}{>}$.  For instance, \texttt{average(a)}, \texttt{std(a)}, \texttt{min(a)}, and \texttt{max(a)} all output the type $a$, whereas \texttt{count(a)} outputs type $count{<}{a}{>}$.  Function evaluations are compatible if their output types are the same.  For instance, \texttt{avg(delay)} is compatible with \texttt{delay} and \texttt{min(delay)}, but not with \texttt{avg(price)} nor \texttt{count(delay)}.

\subsubsection{Expressiveness} 
Expressiveness states that views that {\it could} be composed are allowed by the algebra.   For instance, a bar chart of September \texttt{profits} by \texttt{date} should be composable with a scatterplot of October \texttt{profits} by \texttt{date}.  To achieve this, \sys uses the view's table schema rather than its visual mapping to determine view compatibility (C5).  Of course, there are many compositions that appear sensible to the user, but are disallowed based on our safety heuristics.   For instance, it is natural to compare a chart of \texttt{profits} with a chart of \texttt{losses} because (based on domain knowledge), we know that profits and losses are compatible.   To address this issue, we issue a warning when the user expresses an unsafe composition, and allow the user to override it in a special case: if the dimension attributes match, and the measure attributes are both numeric.  \Cref{s:discussion} discusses additional ways to expand expressiveness.

\section{View Composition Algebra}
\label{s:algebra}

\begin{table}[t]
    \centering \scriptsize
    \begin{tabular}{rl} 
      \textbf{Operator} & \textbf{Description} \\ 
       $\gamma_{A,f(a)}$  & Group by attributes $A$, and compute $f(a)$ for each group \\ 
       $\pi_{e_1\to a_1,\ldots}$  & Compute expressions $e_i$ and rename them as $a_i$.  \\
                        & $T.*$ copies all attributes from input table T. \\
       $\sigma_{p}$  & Filter records using boolean function $p(row)$ \\
       $S\bowtie_{A} T$  & Join $S$ and $T$ rows with the same attribute values in $A$  \\
    \end{tabular}
    \caption{Description of relational algebra operations}
    \label{t:relalg}
\end{table}

\begin{table*}
  \small
  \centering
  \begin{tabular}{llcl c llcl}
    \textbf{Name} &  \textbf{Arity} & \textbf{Notation} & \textbf{Description} && \textbf{Name} &  \textbf{Arity} & \textbf{Notation} & \textbf{Description} \\ 
    Stat Comp & Binary  & $\odot_h(V_1, V_2)$  & Compute difference of matching rows. &&       Extract & Unary & $\downarrow(V,p)$ & Derive subview w/ rows matching predicate $p$.\\
    Union  & Binary & $\cup(V_1, V_2)$ & Superpose or Juxtapose marks.                 &&                                         Explode & Unary & $\Xi_A(V)$ & Facet into small multiples w/ attributes $A$.\\
    Stat Comp & Nary & $\odot_f(\{V_1,\ldots\})$ & Aggregate matching rows from set of views. &&        Lift & Unary & $\uparrow(V)$ & Fit model to view data. \\
    Union & Nary & $\cup(\{V_1,\ldots\})$ & Superpose or Juxtapose marks.& &&&&\\ 
  \end{tabular}
  \caption{Summary of VCA operators.  $V$ denotes a view.}
  \label{t:ops}
\end{table*}

When the user composes two views, what should the output be?  What is the space of possible composition operations?  What views are allowed to be composed together?
A formalism is necessary to ensure that these questions have unambiguous answers.  

This section defines the semantics of a formalism, called View Composition Algebra (\sys), that is designed for visual analytics. We first define view tables as SQL group-by aggregation queries, and then define the major operator semantics\footnote{The remaining definitions can be found in the technical report~\cite{vcatechreport}.}---specifically, the data transforms and visual mappings of the output view given the operator inputs.  \Cref{t:ops} and \Cref{f:algebra_ops} summarize the operators. 
The next section describes interaction designs to express these operators.

\subsection{View Definition}

As described in \Cref{ss:vcaoverview} we model a view as $V_i = R_i(Q_i(D))$.  Let $D(a_1,\ldots,a_n)$ contain $n$ attributes and $A_D = [a_1,\ldots,a_n, a_y]$ denote its schema, where $a_y$ is the measure, and the rest are dimensions.

\subsubsection{Query Transformation}\label{ss:algebra-query}

Most visual analytic systems~\cite{Stolte2000PolarisAS,dice2014,Gray2004DataCA,Chaudhuri1997AnOO,Pahins2017HashedcubesSL} are based on {\it group-by aggregation queries} that filter the input
table using a predicate $p$, group records by a set of attributes $A_{gb}$, and compute an aggregated statistic $f_i(a_y)$:
  \begin{lstlisting}[
    caption={},
    captionpos=b,
    basicstyle=\small\ttfamily,
    frame=,
    mathescape=true,
    escapeinside={@}{!} ]
  $Q_i$ = SELECT @$A_{gb}$!, @$f(a_y)\to y$!  FROM D [WHERE @$p$!] GROUP BY @$A_{gb}$!
  \end{lstlisting}
  \noindent For example, VizQL is a layout algebra for creating faceted, layered charts from relational databases whose statements are translated into group-by aggregation queries~\cite{stolte03thesis}.  \Cref{a:vizql} describes how VizQL statements are expressible as operands in \sys.  

SQL is verbose and difficult to symbolically manipulate, so we will use the equivalent relational algebra statements (\Cref{t:relalg}):

  $$Q_i = \gamma_{A_{gb}, f(a_y)\to y}(\sigma_{p}(D))$$

\noindent $\sigma$ keeps rows that satisfy predicate $p$, $\gamma$ groups rows by a subset of the dimensions $A_{gb}$ and computes  $f(a_y)$ for each group.  

\begin{figure*}
   \centering
   \includegraphics[width=.8\linewidth]{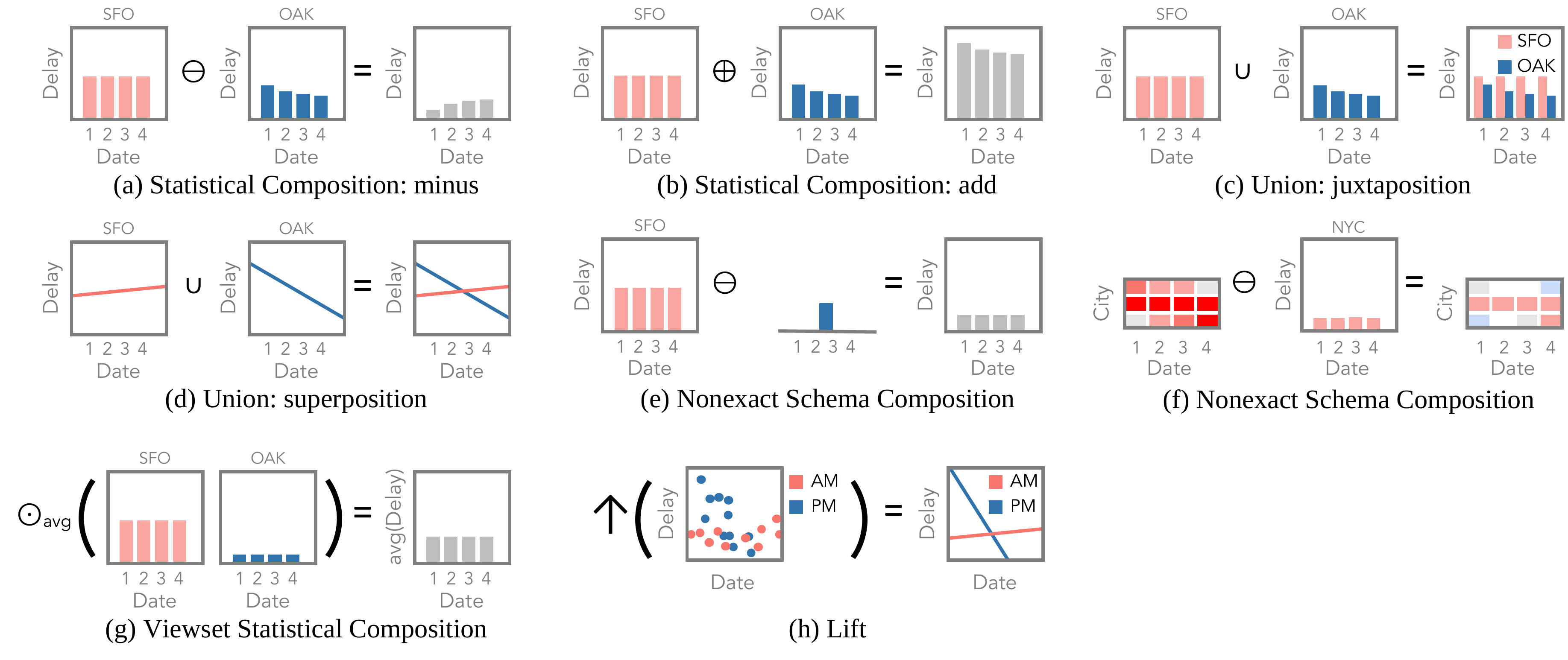}
   \vspace{-1em}
   \caption{Examples of View Composition Algebra operations over views of flight delay statistics.  
     The algebra enables comparison and composition of visualization views.  
     (a) difference between SFO and OAK delay statistics by date,
     (b) the sum of delays by date,
     (c) union of SFO and OAK delay bar marks using juxtaposition,
     (d) union of SFO and OAK delay line marks using superposition,
     (e) difference between 1D SFO delays and a 0D delay scalar,
     (f) difference between a 2D heat map of delays per city, date, and a 1D view of NYC delays,
     (g) average of SFO and OAK delays,
     (h) fit linear models to extrapolate AM and PM delay trends. 
   }
 	\label{f:algebra_ops}
\end{figure*}

\subsubsection{Visual Mapping}

Each output record of $Q_i$ is mapped to one mark.  
Given the view's mark type $K$, $R_i$ maps 
attributes in the query result to visual attributes  (e.g., x, y, color) valid for the mark type $K$. 
Let $A_{Q_i}$ be the attributes in $Q_i$, and $A_K$ be the set of visual attributes for $K$.  $R_i$ is defined as:
$$R_i = \{K\to mark\} \cup \{ a_q\to a_v | a_q\in A_{Q_i} \land a_v\in A_K \}$$
Each visual attribute can be referenced at most once, and not all query attributes need to be mapped.  Let $A_{R_i}$ denote the visual attributes that have been mapped in $R_i$.

Our algebra incorporates Gleicher et al.'s comparison design strategies (juxtaposition, superposition, and explicit encoding)~\cite{gleichercompare} and uses simple default heuristics.  However, sophisticated design strategies are not our main contribution, and the user can always use existing mechanisms to reconfigure the output view's visual mappings or recommend alternative mappings~\cite{Mackinlay2007ShowMA,Moritz2019FormalizingVD,Lin2020DzibanBA}.

\subsection{Binary View Composition}

A binary operator takes views $V_1$ and $V_2$ as input and returns a new view $V^* = R^*\left( Q^* \right) $.  
The statistical operator $\odot_\phi$ derives new statistics from the input views' measures, and
the union operator $\cup_\phi$ merges their data into the same view, and the operator parameter $\phi$ is described below.
\Cref{f:algebra_ops}(a-c) illustrate these binary operators.
Recall that the input views are compatible if the measures are compatible and the set of dimensions are the same.  We will relax the latter requirement in \Cref{sss:schema} to support compositions such as \Cref{f:algebra_ops}(d,e).

\subsubsection{Statistical Composition $\odot$}
\label{ss:algebra_stat}

$V^*=V_1\odot_{op} V_2$ joins rows from $Q_1$ and $Q_2$ and computes a new measure $Q_1.y\ op\ Q_2.y$ from the two views' measures:
\begin{align*}
  Q^* &= \pi_{A_{gb}, Q_1.y\ op\ Q_2.y\to y }\left(Q_1 \fullouterjoin_{A_{gb}} Q_2  \right) \\
  R^* &= R_1
\end{align*}
$Q^*$ first computes the outer join between $Q_1$ and $Q_2$ by matching records from each query 
whose grouping attributes  $A_{gb}$ have the same values.  
An outer join ensures that rows in either table have at least one output row even if there is no match.
Since the input queries, by definition, were grouped on $A_{gb}$, we are guaranteed exactly one output row for each group in $Q_1$ and $Q_2$.  Finally, $\pi$ copies the join attributes, computes $Q_1.y\ op\ Q_2.y$, and renames it as $y$.

$op$ is defined as ``$-$'' by default, however any binary arithmetic function is allowed.
As shorthand, $\oplus$ and $\ominus$ denote $\odot_+$ and $\odot_-$, respectively.
$\odot_{op}$ is symmetric iff $y_1\ op\ y_2 = y_2\ op\ y_1$.

There is one special case where the right operand $V_2$ is handled differently 
than described above.  If a grouping attribute $a\in A_{gb}$ 
only contains a single unique value in $Q_2$'s output,
we remove $a$ from $Q_2$'s grouping attributes.
This is so dimension attributes that do not encode any data variation 
do not affect the semantics of the operator.  
This case is then handled as nonexact schema composition (\Cref{sss:schema}).

\begin{example}
  Riboku wants to know how much worse SFO delays are than OAK.
\Cref{f:algebra_ops}(a) subtracts \texttt{OAK} daily delays from \texttt{SFO} daily delays.  
Since the queries for both views group by \texttt{date, src}, a naive composition
will join the two datasets on $A_{gb}=[date,src]$ and none of the rows will match.
In contrast, dropping \texttt{src} from the join condition is needed compare each \texttt{SFO} day to the correspoding day in \texttt{OAK}.
\end{example}

\subsubsection{Union Composition $\cup$}

Union $V^* = V_1\cup_{qid,a}V_2$ composes the marks from both views into the same output view:
\begin{align*}
  Q^* & = \pi_{*, \hl{qid}}(Q_1) \cup \pi_{*, \hl{qid}}(Q_2) \\
  R^* &= \{qid\to\hl{$a$}\} \cup R_1 \hspace{2em} s.t. \ a \in A_K - A_{R_1}
\end{align*}
Each query $Q_i$ is augmented to track a unique identifier $qid$, so that rows from each query can be distinguished in $Q^*$.  $R^*$ additionally maps $qid$ to a visual attribute $a$ that is not already mapped in $R_1$.

By default, $qid$ is defined based on an internal identifier for the input view. However, it can be any value that uniquely identifies the view.  For instance, if the input views are part of a faceted visualization, then $qid$ may be each view's facet title.   By default, $a$ uses existing perceptual effectiveness orderings~\cite{cleveland1984graphical,Stolte2000PolarisAS} to choose the most effective ordinal visual attribute that is available.

We use heuristics based on the input mark types to decide whether to juxtapose (as in \Cref{f:algebra_ops}(c)) or 
superimpose (as in \Cref{f:algebra_ops}(d)) the marks.  If the mark fills the area between the measure value and 0 (e.g., bar charts, area charts), then the marks in the output view are juxtaposed to avoid overlap.
Otherwise, the marks (e.g., line and scatterplot) are superimposed.
The user can always reconfigure the output view based on their own preferences.

\subsection{Statistical Composition with Nonexact Schemas}
\label{sss:schema}

We now relax the safety rules for statistical composition $\odot_{op}$ for cases like \Cref{f:algebra_ops}(e,f) where the two views do not have identical query schemas, but the grouping attributes $A_{gb}^1$ in $Q_1$ are a strict super set of the grouping attributes $A_{gb}^2$ in $Q_2$.  In these cases, each row in $Q_2$ potentially matches many rows in $Q_1$, and so we perform a {\it left} outer join between $Q_1$ and $Q_2$ to match all of them.    This ensures that all rows in $Q_1$ are preserved, but rows in $Q_2$ without a join match are not in the output.  This is because $V_1$ is transformed by matching rows in $V_2$ but not vice versa.  The result also preserves all grouping attributes in $Q_1$, and renders the result using $V_1$'s visual encoding.  The semantics depends on how $A^2_{gb}$ is defined:

\begin{example}
  Shin wants to compare SFO delays with OAK's delay yesterday (a scalar).
  \Cref{f:algebra_ops}(e) composes a 1D view of SFO delays ($V_1$) with a 0D view ($V_2$) 
  consisting of a single delay constant.
  The scalar is subtracted from each day's delay in the SFO chart.
\end{example}

\begin{example}
  Ouki is analyzing daily delays across cities in the US, and wants to compare them with NYC.
  \Cref{f:algebra_ops}(f) composes a 2D heat map of delays by city$\times$date ($V_1$)
  with a 1D bar chart of NYC delays by date ($V_2$).  
  The resulting heatmaps removes OAK's delays from each city's delays.
\end{example}

\subsection{View Decomposition}\label{ss:decompose}

It is often useful to decompose components of a view (values, marks, groups) into one or more stand-alone views that contain a subset of the input view's rows and attributes.  The new views can then be used as operands for further composition, or manipulated for further analysis.  To this end, we define 2 decomposition operators based on interactions often used during view manipulation and coordination.   

The {\it extract} operator $\downarrow(V, p) = R(\sigma_p(Q))$ creates a single new view based on a subset of rows in $V$ that match predicate $p$.  If $p$ is not specified, it defaults to $true$, which is equivalent to creating a copy of $V$.  $\downarrow$ can be used to create a standalone view that renders a selected subset of marks, similar to coordinated overview-detail visualizations.   $\downarrow$ can also be used to e.g., extract marks from different views and then juxtapose them using $\cup$.  

The {\it explode} operator $\Xi_{A_e}(V)$ generates a set of views (a viewset), with one view for each group defined by the attributes $A_e$.   Explode is similar to facetting, which is traditionally used when creating new views from raw data.  In contrast, $\Xi$ is directly applied to existing views, including views derived from previous manipulations or compositions.  
We define this formally in the tech report~\cite{vcatechreport}.

\subsection{Composing Viewsets}\label{ss:viewsets}

We now extend composition to a {\it viewset} containing $n$ views 
$\mathbb{V} = \{V_1,\ldots,V_n\}$.  
Following our safety rules, all views in $\mathbb{V}$ have identical schemas, and
their queries  have the same grouping attributes $A_{gb}$.  

\subsubsection{Viewset Statistical Composition $\odot$}

When statistically composing a viewset $V^* = \odot_{f^*}(\mathbb{V})$, 
the purpose is to derive new statistics from the underlying data of each of the views,
rather than compare any individual pair of views in the viewset.
For this reason, we first union the {\it preaggregated} data from each query, and then
{\it reaggregate} the union.  
These semantics differ from binary statistical composition, which composes the aggregated measures from each input query.

We first rewrite each $Q_i$ as $\gamma_{A_{gb}, f(a)\to y}(q_i)$, 
where $q_i$ represents any filtering, join, and projection
operations that $Q_i$ performs prior to grouping and aggregation ($\gamma$).
This does not change the semantics of $Q_i$, but allows us to refer to $q_i$ in the definition above.
\begin{align*}
  Q^* = \gamma_{A_{gb}, \hl{$f^*(a)$}\to y}\left(q_1 \cup \ldots \cup q_n  \right) 
  \hspace{3em}
  R^* = R_1
\end{align*}
$Q^*$ first unions all $q_i$ before grouping by $A_{gb}$ and 
computing the aggregation function \hl{$f^*\left(  \right)$}.  
We do this to ensure that aggregate statistics are always computed over the input data, 
and so users do not inadvertantly compute e.g., averages of averages, 
which can easily lead to misinterpretation. 
Common aggregations include \texttt{min}, \texttt{max}, \texttt{avg}, \texttt{std}, and \texttt{count}, however any 
numeric aggregation function (e.g., $\frac{avg(delay)}{max(delay)}$ is acceptable.
We use $f^*\left(  \right) $ to distinguish the parameter from the non-aggregation function 
$h\left(  \right)$ used in binary composition and the aggregation $f\left( \right)$ in the input
queries.

\begin{example}
   Kanki wants a sense of the typical daily delays in the Bay Area (SFO and OAK airports).  \Cref{f:algebra_ops}(g) computes the average of a viewset containing SFO and OAK delay charts.  Their delays are grouped by Date and averaged to produce the output average daily delays.  
\end{example}

We also generalize the special case in binary statistical composition to every view in the viewset.   
If a grouping attribute $a$ has
a unique value in {\it every} view's query result, it is removed from 
$A_{gb}$ when computing the group by in $Q^*$.
For instance, if the user highlights a set of marks in a chart, and wants to summarize
their measures, then each mark is treated like a separate view, and their grouping attributes are
all dropped (since there is a single group in each ``view'').
If this is not done, the output in the example would be the same set of marks---effectively a noop.
We illustrate this interaction in \Cref{ss:iact_nary}.

\subsubsection{Additional Viewset Operations}

The union operator $V^*=\cup_{qid,a_t}(\mathbb{V})$ is a direct generalization of 
binary union.  It adds \hl{$qid$} to each query in the viewset, unions the queries, 
and maps \hl{$qid$}  to an available visual attribute.
Binary operators can also be applied to viewsets by applying the operator to every combination of views in the input viewsets.

\subsection{Learning and Composing Model Views $\uparrow$}

Two views may not neatly compose because the {\it values} of the grouping attributes do not match.  
For instance, suppose $V_1$ renders delays on even days while $V_2$ renders  delays on odd days.  
Although $V_1\ominus V_2$ is safe, the resulting view
is uninformative because there are no matching dates in the join.
For this reason, the unary {\it lift} operator $\uparrow_{M,A_d, A_c}(V)$ fits a model $M$ to the data in view $V$ and returns a {\it model view}:  
\begin{align*}
  Q^* &= \gamma_{\hl{$A_c$}, train_{\hl{$M$}}(\hl{$A_d$},y)}\left(Q \right) \\
  R^* &= \{a\to a_v |  a\to a_v\in R \land a\in \hl{$A_c$}\cup \hl{$A_d$} \}\hspace{2em} s.t. \ \hl{$A_d, A_c$} \subseteq A_{gb}
\end{align*}
In the above, $Q^*$ groups the output of $Q$ by $A_c$, and for each group, trains a model $M$ that uses $A_d$ as features to predict the measure $y$.  In effect, the operator trains the model $M(A_d, y | A_c)$.
$R^*$ only keeps mappings in $R$ related to the features $A_d$ or conditioned attributes $A_c$.
We require that $A_d$ and $A_c$ are part of $V$'s grouping attributes, and $A_d$ contains only quantitative attributes that can be used as numeric features in the model.

\begin{example}
 Heki  wants to understand the overall delay trends in the mornings and afternoons.   In \Cref{f:algebra_ops}(h), $\uparrow_{linear,[date],[ampm]}(V)$ fits two linear regression models, one for AM and one for PM.  Each model uses \texttt{date} and an intercept term to predict the average delay.  
\end{example}

Models are a compact representation of an infinite-sized relation, however
 charts render finite sets of rows.
In order to render a model view, we sample the domain of $A_d$ and predict
the $y$ value for each sample in each group as defined by $A_c$.  
The results are used as the ``query result'' and rendered as normal.
As a heuristic, we sample $20$ equi-distant values from the domain of each attribute in $A_d$; if there are many attributes in $A_d$, we scale down the samples per attribute so the total number of samples does not exceed $1000$.

\subsubsection{Composing a View and Model View}

A major use case for model views is to aid comparisons
between views that would otherwise not share join values.  
We define the composition  $V_1\circ V_M$ 
of a view $V_1$ and model view $V_M$ as follows, where $\circ$ is any binary composition operator and the usual safety rules apply:
\begin{align*}
  Q_M' &=  \pi_{A_{gb}^1, M(Q_1.A_d)\to y}\left( Q_1 \bowtie_{A_c} Q_M \right) \\
  V_M' &= R_M(Q_M')\\
  V^* &= V_1 \circ V_M'
\end{align*}
At a high level, we use the models in $V_M$ to predict the $y$ value for each
row in $Q_1$.  To do so, we need to match the appropriate model $M$ for each $Q_1$ row.  We do this using an inner join on $A_c$, which emits a join result only if there is a match.    This is to avoid the case when a model has not been trained for the specific $A_c$ values for a row in $Q_1$; we emit a warning with the number of such rows.   Each join result projects the grouping attributes $A_{gb}^1$ from $Q_1$ along with the predicted $y$ value.     If $\circ$ is symmetric, then $V_M\circ V_1 = V_1\circ V_M$.

\subsubsection{Composing Two Model Views}

In order to compose two model views, we follow the same procedure as rendering a model view.  We sample from $A_d$, and use the samples to generate predictions from both model views.  The results are treated as query results and the normal composition operator is applied.

\subsection{Properties of the Algebra}

The operators are closed---operator outputs can be inputs to other operators---except in one case.
The n-ary statistical composition operator assumes that the input views' queries in canonical form  $\gamma_{A, f()}(q)$, where $q$ does not contain any aggregation operators.  All operators return views with canonical queries, except for binary statistical composition $\ominus(V_1, V_2)$, whose output query structure is $\pi(Q_1\fullouterjoin Q_2)$.   In this case, aggregation happens before the join ($\fullouterjoin$) and projection ($\pi$), and in general cannot be rewritten into canonical form.

\subsection{Compatibility With VizQL}\label{a:vizql}

VizQL~\cite{stolte03thesis} is a visual specification that describes table-based visualization layouts of multi-dimensional data.  The key elemnt of the specification is a table algebra that is defined over the ordinal and quantitative data attributes in the input database.  
Users specify table algebra expressions to organize the layout along the x- and y-axes to define small multiple charts.

\begin{figure}
\centering
\includegraphics[width=\columnwidth]{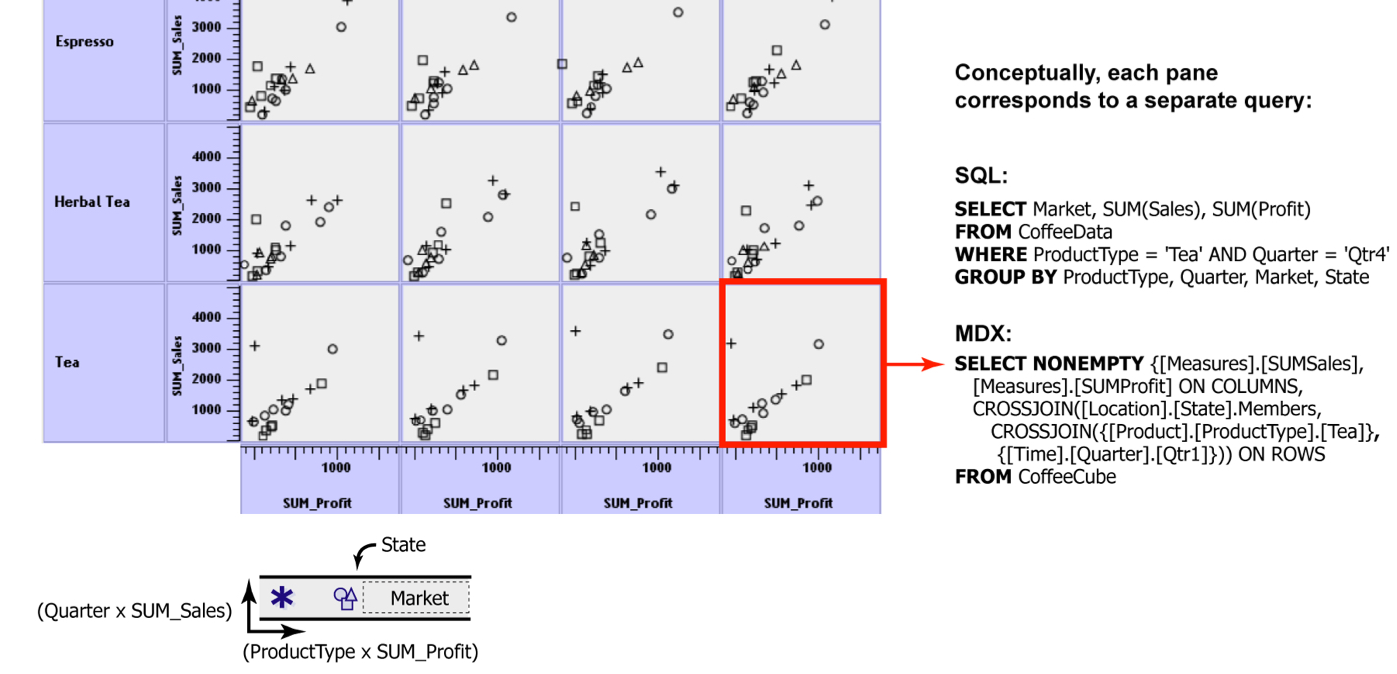}
\vspace{-.2in}
\caption{VizQL specification plots total profits (x-axis) against total sales (y-axis) for each market (shape of mark) and state.  The layout is faceted by product type and quarter along the x and y dimensions.}
\label{f:vizql}
\end{figure}

\begin{example}
  The lower left in \Cref{f:vizql} is the VizQL specification that renders the faceted visualization.
  The x- and y-axis table structures are defined by the table algebra statements \texttt{Product$\times$SUM(profit)} and \texttt{Quarter$\times$SUM(sales)}, respectively.  The first statement creates a separate pane for each unique \texttt{Product} value that renders the corresponding subset of the input table.
  The specification further maps \texttt{Market} to the mark shape, and groups records by \texttt{State}.   Thus, each mark reports statistics for a given product type in a specific state's market, in a particular quarter.
\end{example}

Despite the flexible table algebra to specify visualizations, VizQL directly translates into simple SQL group-by aggregation queries that are in the same form that \sys supports. The algorithm on page 68 in Stolte et al.~\cite{stolte03thesis} describes the 4 steps that define the query's expressions in the \texttt{SELECT} clause, the filters in the \texttt{WHERE} clause, and the fields to \texttt{GROUP BY}.    For example, the step to construct the query's \texttt{GROUP BY} clause is:
\begin{enumerate}[]
  \setlength\itemsep{0em}

  \item Add categorical fields from the x- and y-axis p-tuples,
  \item Add categorical fields form all encodings, and
  \item Add fields from the ``Group'' to the GROUP BY
\end{enumerate}

\noindent A notable aspect of the VizQL to SQL translation is that it only refers to the fields that are referenced in the p-tuple expressions, and not the algebraic operations over the fields.  This is because the algebra expressions are for {\it rendering and layout}, and not for data processing.

\begin{example}
  \Cref{f:vizql} is easily translated into  a set of group-by aggregation query following the algorithm in Stolte et al.~\cite{stolte03thesis}.  Each pane in the visualization encodes a query whose group by attributes are $A=Product, Quarter, Market, State$.  For instance, the highlighted pane in the lower right is a view $V=R(Q)$ defined as:
  \begin{align*}
    Q &= \gamma_{A, sum(sales)\to x, sum(profit)\to y}(\sigma_{product=tea, quarter=4}(T)\\
    R &= \{ point\to mark, x\to x, y\to y, Market\to shape  \}
  \end{align*}
\end{example}

\noindent For these reasons, VizQL specifications are compatible with \sys.

\begin{figure*}[t]
  \centering
  \includegraphics[width=\textwidth]{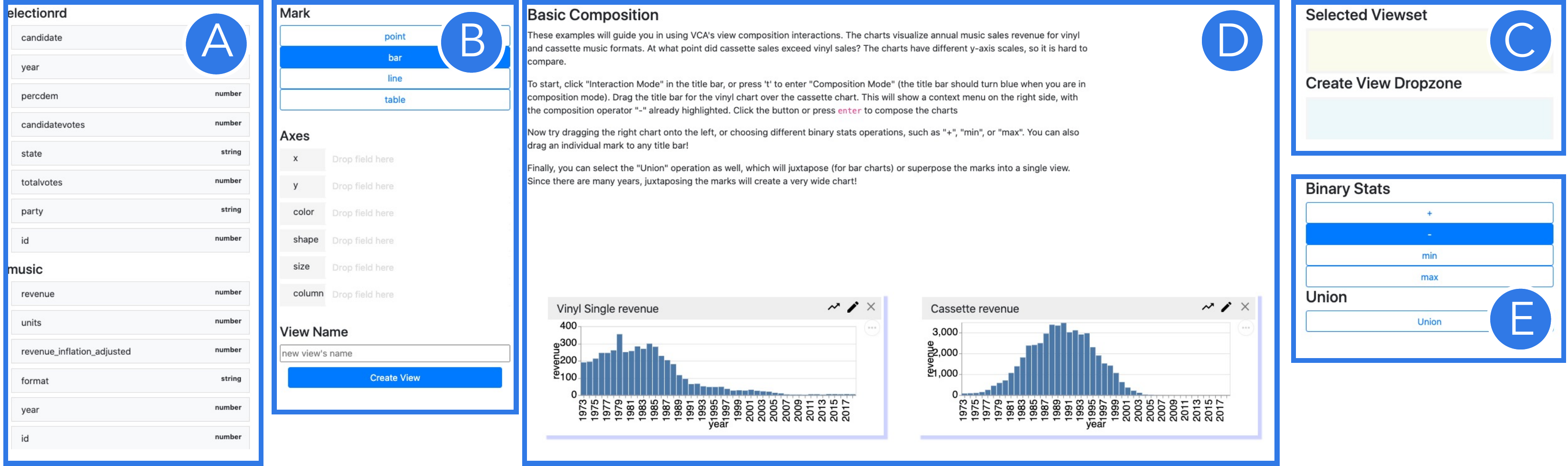}
  \caption{Our prototype interface augments Tableau-like view creation functionality with view composition.}
  \label{f:screenshot}
\end{figure*}

\section{Interaction Design and Implementation}\label{s:interaction}

This section describes one possible interaction design for comparison.
The design seeks to easily express common cases via direct manipulation and sensible defaults, and support customization through context menus.  
Users can 1) compose views by dragging one view onto another, where the destination and dragged views are the left and right operand, respectively; 2) decompose components of a view into new views; 3) define viewsets by selecting one or more views; and 4) lift a view into a model view.  
We showcase these interactions using a Tableau-like interface for creating, filtering, and composing charts.
We include screenshots of these direct manipulation gestures for most examples, and use simplified diagrams when it is more clear. 
The examples in this section will primarily be based on the \texttt{cars} dataset, where the user is analyzing relationships between cars with different numbers of carburetors (\texttt{carb}), cylinders (\texttt{cyl}), horsepower (\texttt{hp}), and miles per gallon (\texttt{mpg}).

\subsection{Interface and Implementation Overview}

We have implemented a prototype interface that augments a Tableau-like view creation functionality with view composition interactions (\Cref{f:screenshot}).  \bluecircled{A} and \bluecircled{B} are panes similar to Tableau~\cite{Stolte2000PolarisAS} and Voyager~\cite{voyager}: \bluecircled{A} lists the database tables and attributes that can be dragged to visual attribute shelves \bluecircled{B} to create new views.      \bluecircled{D} renders the set of visualizations that the user can compose, as described in this section.  \bluecircled{C} are two dropzones: the {\it Selected Viewset} dropzone lists the views in the user's current viewset and can be used as a proxy for viewset composition, and the {\it Create View Dropzone} is used for view decomposition.    Finally, \bluecircled{D} renders the context menus for composition operations.    To disambiguate between view manipulation and compositions interactions, the user first toggles ``Composition Mode'' before performing \sys operations.

The prototype uses a library of visualization implementations.  It renders statistical charts using Vega-lite~\cite{Satyanarayan2017VegaLiteAG}, tables using browser native \texttt{table} tags, and maps using the Leafletjs library\footnote{\url{http://Leafletjs.com}}.
The prototype executes queries on a client-side sql.js database (this disables the {\it lift} operator, which requires a database that supports ML) as well as a remote database connection (we use a PostgreSQL database with the Madlib~\cite{Hellerstein2012TheMA} machine learning library).  
The prototype is built on top of a standalone \sys.js Javascript library, which implements internal SQL query and visual encoding representations, \sys operators.

\subsection{Specifying View Operands}

New visualization types are integrated by implementing a render function that takes as input the view's query result and visual mapping specification.  The visualization is rendered with a title bar that users can drag to use the entire view as an operand.  Subcomponents of the view are supported by defining interaction handlers to generate the corresponding queries and visual mappings.  This subsection illustrates this for charts, table, and map visualizations.

\subsubsection{Chart Visualizations}

\begin{figure}[H]
  \centering
  \vspace{-1em}
  \includegraphics[width=\columnwidth]{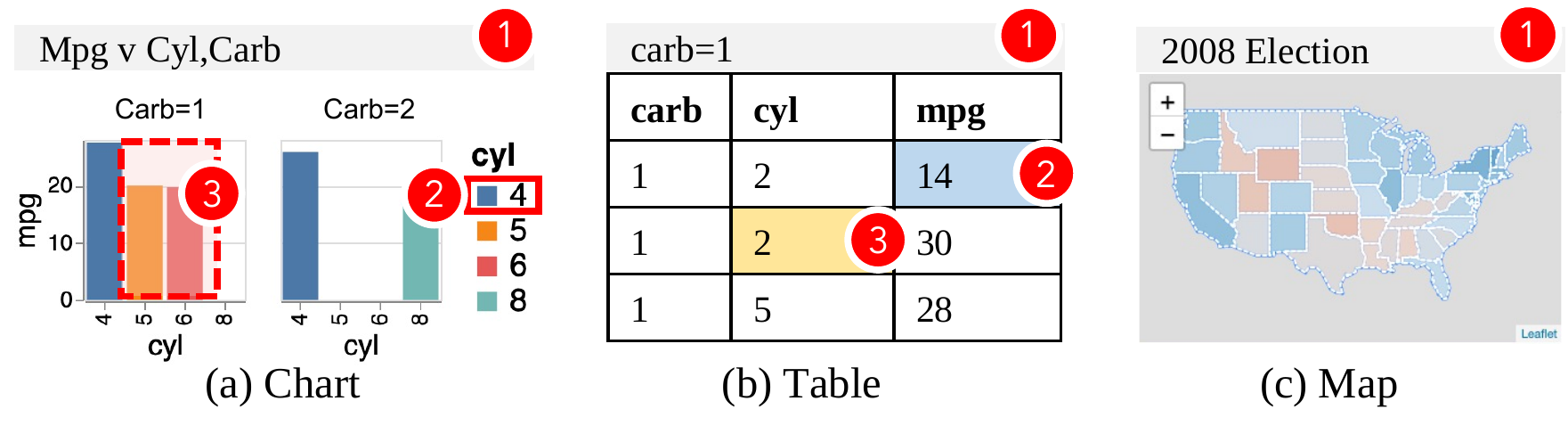}
  \vspace{-1em}
  \caption{Each component of a rendered view can be interactively chosen as an operand for view composition.  
  This figure classify 3 operand types: the entire view, grouping attribute values, and marks. }
  \label{f:iact_viewtypes}
\end{figure}

We define four types of chart operands---the chart, a legend label, marks, and a value--that users can specify.  Given an existing view (e.g., \Cref{f:iact_viewtypes}(a)), we show how a user specifies a view component as an operand using a bar chart and table.  The formal definitions are in the technical report~\cite{vcatechreport}.

Our implementation uses Vega-lite to render the chart, and listens to the Vega-lite signals to turn interval selections into a new view.  We also instrument legend label dom elements using custom event handlers.

\stitle{\circled{1} Entire View:} the most straightforward is to select the chart's title bar to use the entire view as the operand.  

\stitle{\circled{2} Labels in Legend:} 
the user can drag a label in the legend to pick a subset of rows matching the selected label.  Following the rules in \Cref{ss:algebra_stat}, the attribute that the label corresponds to is dropped from the subset of rows as well as the visual mapping.  

\begin{example}
  Ouhan in \Cref{f:iact_binaryop}(e) drags the \texttt{carb=1} label onto the \texttt{carb=5} chart and chooses $\ominus$.  Since the value of \texttt{carb} in the dragged view is unique, it is dropped so that the join condition is on \texttt{cyl}. Otherwise, the join condition would require matches on cylinder {\it and} carburetors, which would not contain any matches.    If Ouhan chose the $\cup$ operator instead, \texttt{carb} would be preserved.
\end{example}

\stitle{\circled{3} Marks:} the user can drag-select a set of marks to define a view with the selected rows and same visual encodings.  

\stitle{Constant Value:}  The user can specify a constant value via a form. It corresponds to a view with no mappings.  Since the value is not bound to an attribute in the database, it is compatible with any measure attribute.  Also, since it does not define mappings, the constant value is restricted to the right operand of a binary operator.

\begin{figure}
  \centering
  \includegraphics[width=.7\columnwidth]{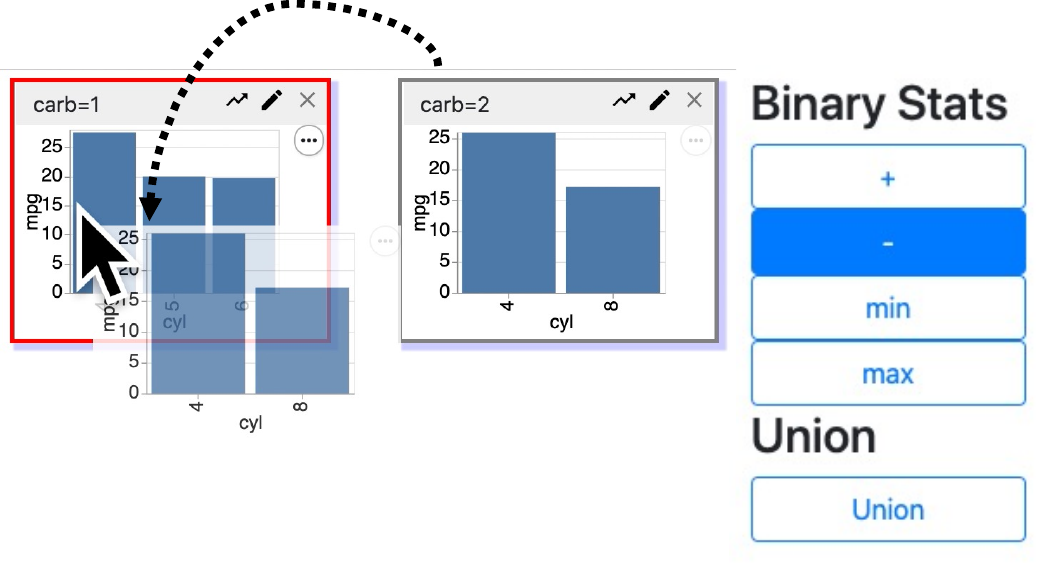}
  \vspace{-1em}
  \caption{Binary composition: context menus specifies the operator and its parameters.  \texttt{<enter>} accepts highlighted defaults. }
  \label{f:iact_menu_binary}
\end{figure}

\subsubsection{Table and Map Visualizations}

Tables support 3 operand types: table, measure value, and set of rows.  \Cref{f:iact_viewtypes}(b) contains two dimension and one measure attributes.  In addition to dragging the title bar to use the table as an operand, the user can \circled{2} drag a measure cell (e.g., 14 in blue) as a view containing one attribute (\texttt{mpg}) with one value (14) and \circled{3} drag a dimension value (e.g., \texttt{2}) to select all rows with that dimension value.   Similar to the legend label, the selected dimension attribute is dropped from the query and visual mapping if the operand is used in binary composition.   

We also integrated Leaflet-based map visualizations to support using the full map as an operand  (\Cref{f:iact_viewtypes}(c)). The integration took ${<}60$ lines of code, mainly for setting up the styling.   Using subcomponents (such as selecting a State) as operands can be supported by further using Leaflet's event API, which we leave to future work.

\subsection{Binary View Composition}

\begin{figure}
  \centering
  \vspace{-1em}
  \includegraphics[width=.8\columnwidth]{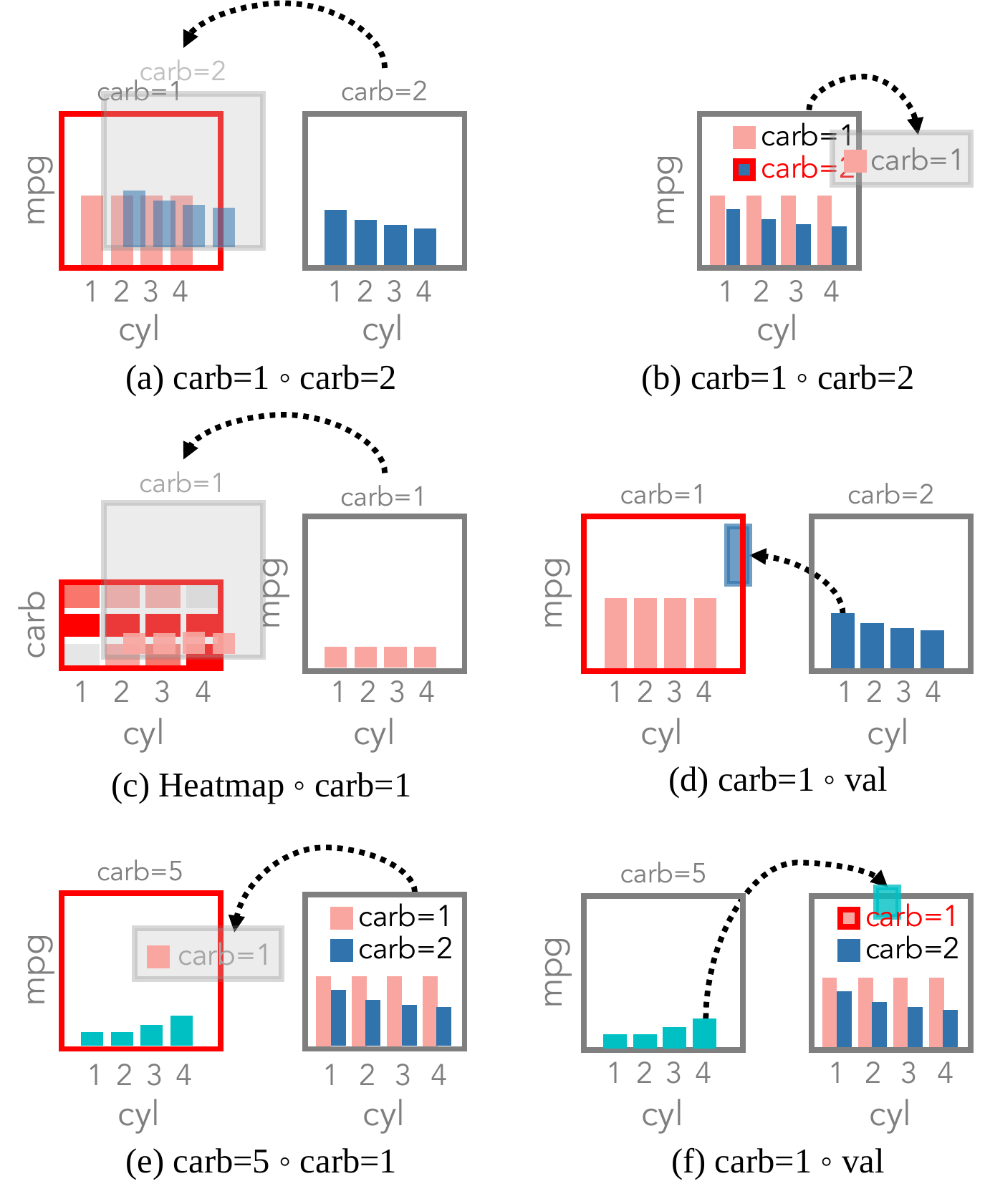}
  \vspace{-1em}
  \caption{Diagrams of drag interactions to specify binary operands.  
    The source and destination correspond to the right and left operands.
    (a) drags the \texttt{carb=2} view onto the \texttt{carb=1} view,
    (b) drags the \texttt{carb=1} group label onto the \texttt{carb=2} label,
    (c) drags the \texttt{carb=1} view into the 2D heatmap,
    (d) drags a value onto the \texttt{carb=1} view,
    (e) drags the \texttt{carb=1} group label onto the \texttt{carb=5} view, and
    (f) drags a value onto the \texttt{carb=1} group label.
  }
  \label{f:iact_binaryop}
\end{figure}

The user specifies a binary composition by dragging a source view (the right operand)  onto the destination view (left operand).  For instance, \Cref{f:iact_binaryop}(a) drags the \texttt{carb=2} view onto the \texttt{carb=1} view to express $carb=2{\circ}carb=1$.  Once the user has specified the source view as described above, the interface highlights the charts that are safe destinations.  The user can drag the right operand onto the body of the chart, a legend label, or an individual mark; note however, that marks may overlap and be difficult to drag onto.  Sets of marks are supported turning them into standalone charts (via view decomposition, described below). \Cref{f:iact_binaryop} depicts examples of drag interactions and their corresponding algebra statements.   The destination views are highlighted in \red{red} and the interaction is depicted as the dashed arrow.

The user then specifies the operator ($\odot$ or $\cup$) and its parameters using a context menu (\Cref{f:iact_menu_binary}). Statistical composition lists common arithmetic functions.  By default, the operator is $\ominus$, and the user can press \texttt{<enter>} to accept the highlighted default.     If the views are not safe to compose, the user is warned and can optionally override the warning.  In this way, users can use their domain knowledge to judge when exceptions to our safety heuristics should be made.

\begin{figure}
  \centering
  \vspace{-1em}
  \includegraphics[width=.85\columnwidth]{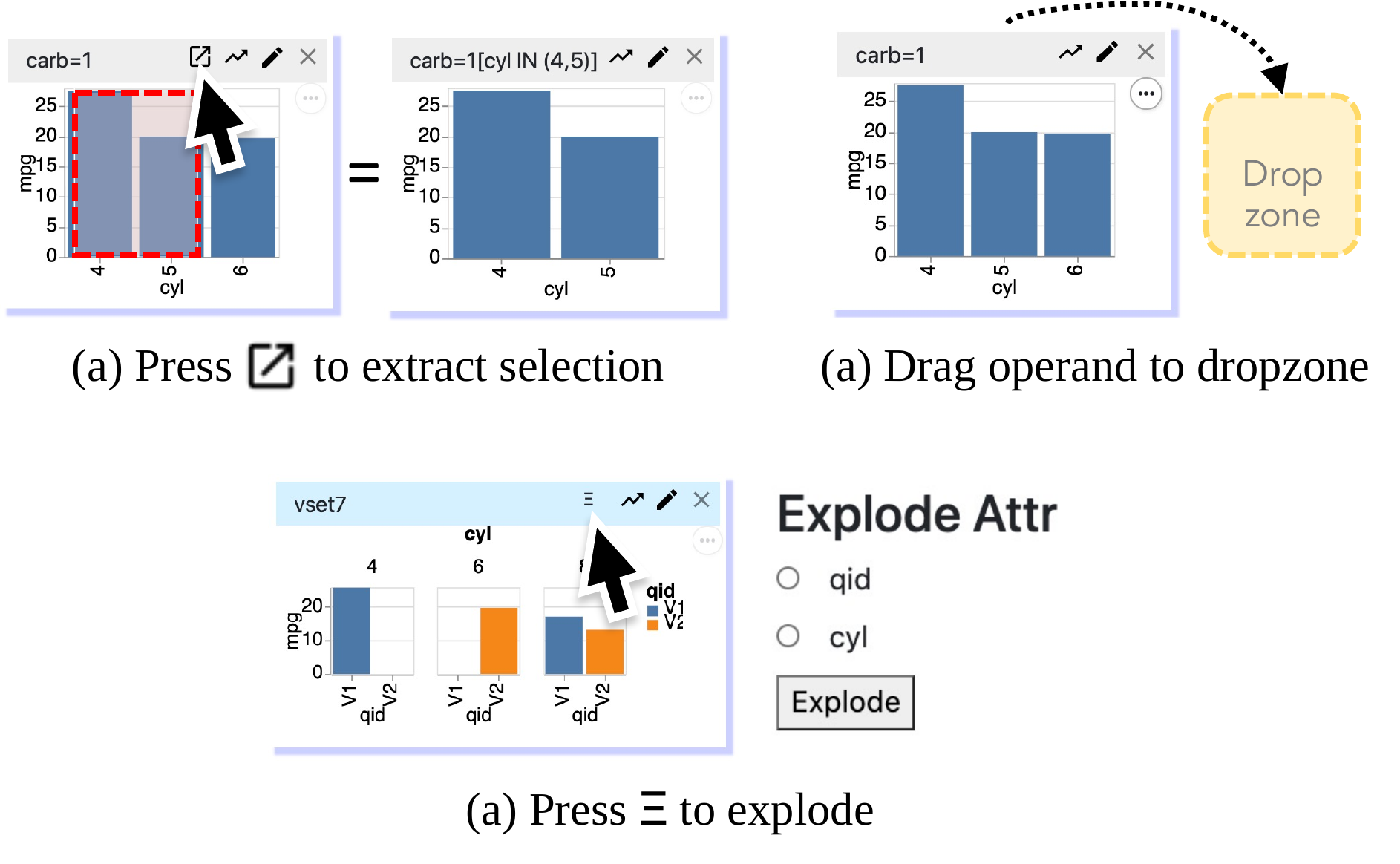}
  \vspace{-1em}
  \caption{Decomposition interactions. (a) Extract a selected set of marks in a view, (b) extract an entire view, mark, or legend label, and (c) explode a view.}
  \label{f:iact_decompose}
\end{figure}

\subsection{View Decomposition}
An operand is a view, and the user can create a standalone view by dragging it onto the ``create view'' dropzone.  This is useful when the user wants to compare an individual or set of marks---they can select the marks and drag them to the dropzone, which creates a new view containing only those marks.  The user can then manipulate, interact with, or compose the new view directly.

\begin{example}
  In \Cref{f:iact_decompose}(a), Mouten selects cylinders 4 and 5 and presses \includegraphics[height=1em]{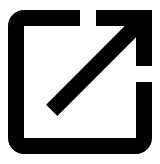}.  This generates a predicate $p: cyl\in[4,5]$ that is passed to extract $\downarrow$ to create a new view.  In \Cref{f:iact_decompose}(b), Mouten drags the view (or a legend label or mark) onto a drop zone to extract it into a new view.   In \Cref{f:iact_decompose}(c),  Mouten clicks {\bf $\Xi$} in the title bar to open the context menu and choose the feature and conditioned attributes. 
\end{example}

\subsection{Nary Viewset Composition}
\label{ss:iact_nary}

\begin{figure}
  \centering
  \includegraphics[width=\columnwidth]{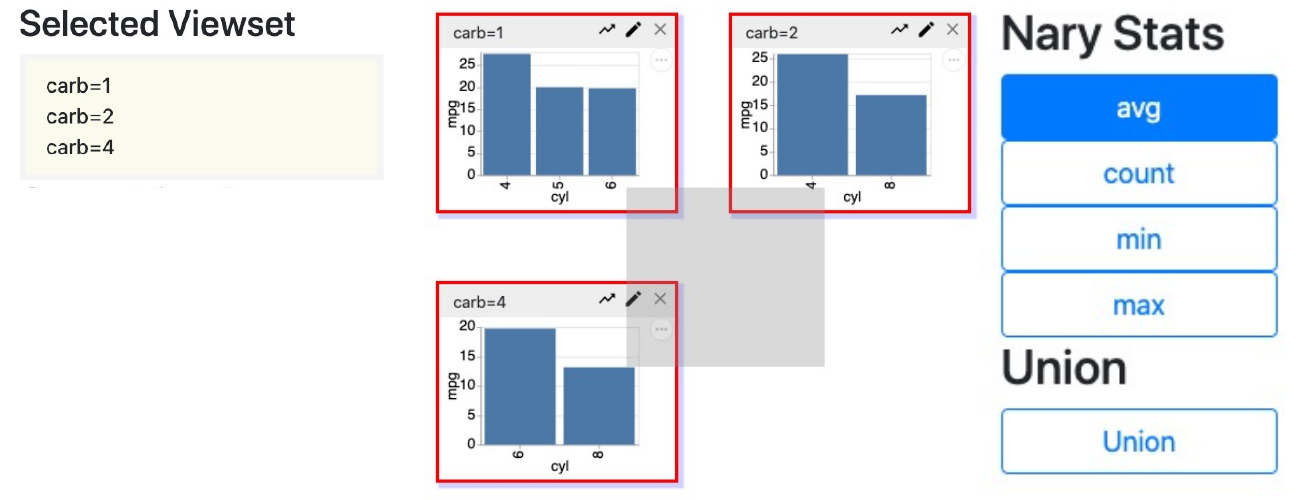}
  \vspace{-1em}
  \caption{Nary composition: user selects a set of views, and uses context menus.   \texttt{<enter>} accepts highlighted defaults. }
  \label{f:iact_menu_nary}
\end{figure}

The user specifies a viewset by brushing over a set of views (\Cref{f:iact_menu_nary}).  
This also updates the ``Selected Viewset'' dropzone, which is a representation the user can use as a proxy to compose the viewset with another view.
The user uses the context menu to specify the statistical aggregation function or the union operator.

\subsection{Model View Composition}

The user lifts a view by clicking on \includegraphics[height=1em]{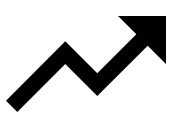} in the title bar.  This brings up the context menu where the user can choose the model (linear, logistic, or generalized linear model) and the feature and conditioned attributes.  
The feature attributes are used to predict the view's measure, 
and a separate model is fit for each unique group of rows as defined by the conditioned attributes.
Model training and inference is handled by the Apache Madlib PostgreSQL extension~\cite{Hellerstein2012TheMA}, which is a library of machine learning data types and functions.

\begin{figure}
  \centering
  \includegraphics[width=.65\columnwidth]{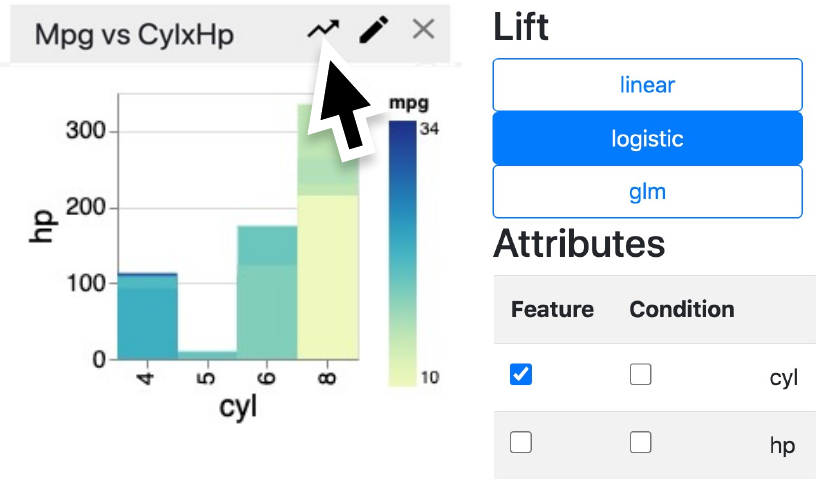}
  \vspace{-1em}
  \caption{Context menus for lift operation. The user clicks \protect\includegraphics[height=1em]{figs/icon2.png} in the title bar and chooses the model, features, and conditioning attributes.    \texttt{<enter>} accepts highlighted defaults. }
  \label{f:iact_menu_lift}
\end{figure}

\begin{figure*}
  \centering
  \includegraphics[width=0.8\textwidth]{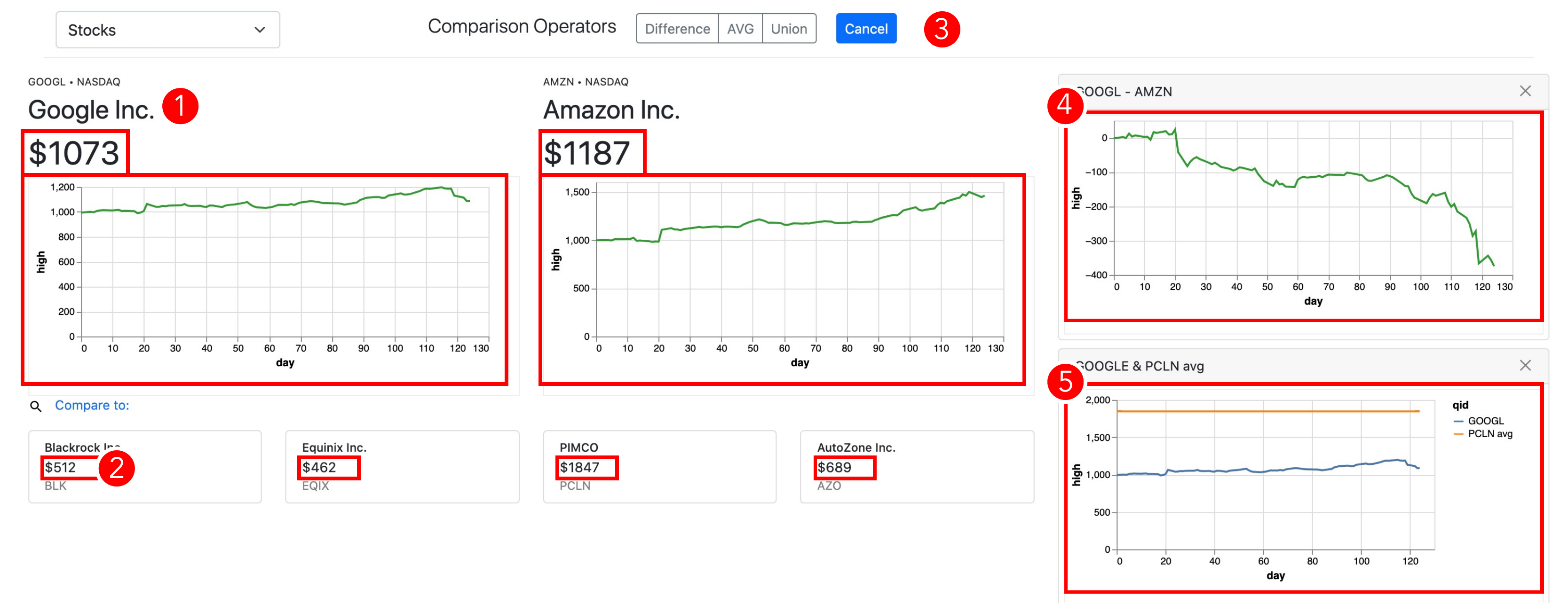}
  \caption{
    Stock analysis application based on Google Finance, showing recent stock prices for Google and Amazon,
    and summaries of four related companies.  The user can click-drag to compare any
  data in the interface (highlighted in red).  Comparison results are listed vertically on the right.}
  \label{f:google_demo}
\end{figure*}

\section{Examples}\label{s:eval}

\sys enables direct comparisons in tasks that require summarizing and/or comparing data in one or more views, and the examples so far have illustrated basic usage of individual operators.  This section presents novel comparison examples that are difficult or impossible to express in existing visual analysi systems, such as comparing across visual encodings, multi-step compositions, and combining \sys with view manipulation interactions.   It ends with a case study based on Google Finance.

\subsection{Comparing Across Visual Encodings}

Even in bar charts, design factors such as alignment, the distance between bars, and their values can effect visual comparison judgements between pairs of bars.   It can be even more challenging to compare values that are encoded using different mark types or visual attributes.  \Cref{f:tableau} was an example where the user wants to compare statistics encoded across text labels, lines, and bars.
\sys's {\it design independence} easily supports this.

\begin{figure}[H]
\centering
\includegraphics[width=.75\columnwidth]{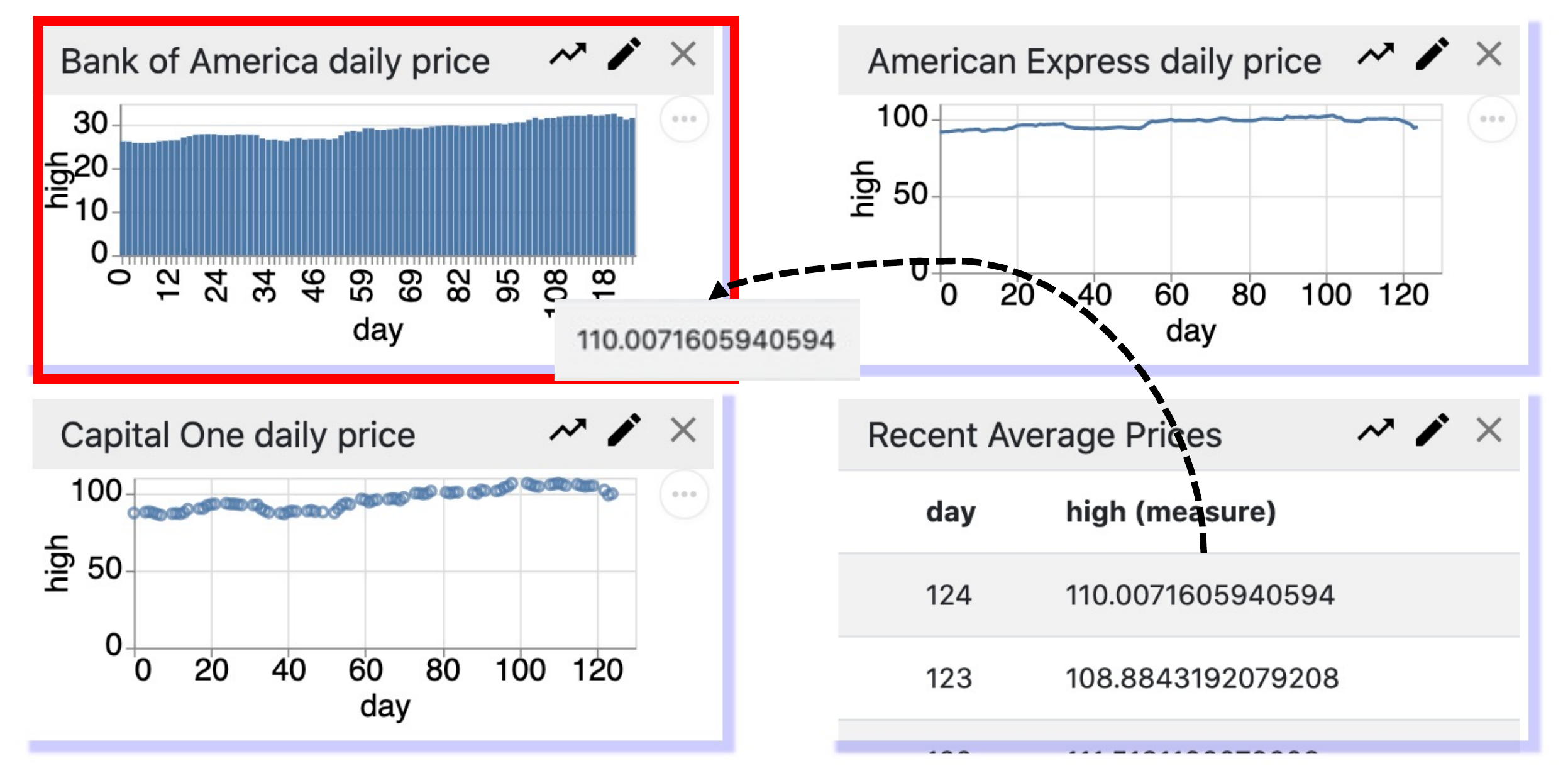}
\vspace{-.2in}
\caption{\sys can compare data rendered using different visual encodings.}
\label{f:exp_crossenc}
\end{figure}

\begin{example}
\Cref{f:exp_crossenc} illustrates four views that use different visual encodings.  The bar chart, line chart, and scatter plot render daily stock prices for Bank of America, American Express, and Capital One, and the table renders recent daily average prices as text.   All of the views are composable because their underlying schemas can be matched.  For instance, the user in the figure is dragging the average high price across all S\&P 500 stocks on day 124 onto the Bank of America bar chart to compare their statistics.  
\end{example}

\subsection{Multi-step Composition Patterns}

A benefit of \sys's algebraic formulation is that (most) composition outputs can be composition inputs, to form multi-step composition-based analyses.  
We now present three examples where Alice is analyzing fuel efficiency (mpg) by cylinder (cyl) for cars with 1, 2, or 5 carburetors (carb), and wants to know how 
different carburetors differ from the overall fuel efficiency.
Each example differs from the previous example by one operation, however they result in very different analyses.  
For conciseness, the figures use the algebraic notation, however 
these patterns are expressible using interaction gestures.  

\begin{figure}[H]
  \centering
  \includegraphics[width=\columnwidth]{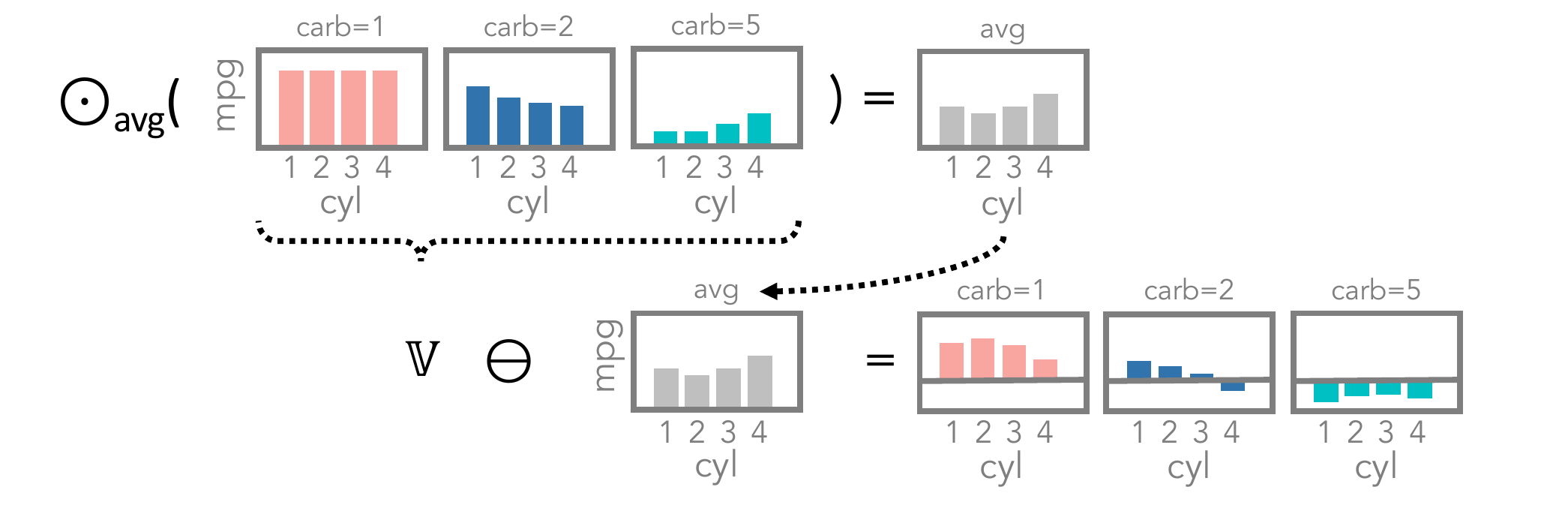}
  \vspace{-.2in}
  \caption{Alice computes the average miles per gallon (mpg) by cylinder across the different carburetors, then removes the averages from each carburetor plot. }
  \label{f:iact_avgminus}
\end{figure}

\begin{example}
  Alice wants to understand how the number of carburetors (1, 2, 5) 
  affects fuel efficiency compared to the overall for different numbers of cylinders.
  \Cref{f:iact_avgminus} shows the steps.   
  The user first creates a viewset $\mathbb{V}$ by selecting the three charts,
  and uses $\odot_{avg}$ to group mpg for each cylinder, and computes the average mpg for each cylinder.
  She then drags the average chart over the viewset and picks $\ominus$ to see the comparison.
\end{example}

\begin{figure}[H]
  \centering
  \vspace{-1em}
  \includegraphics[width=\columnwidth]{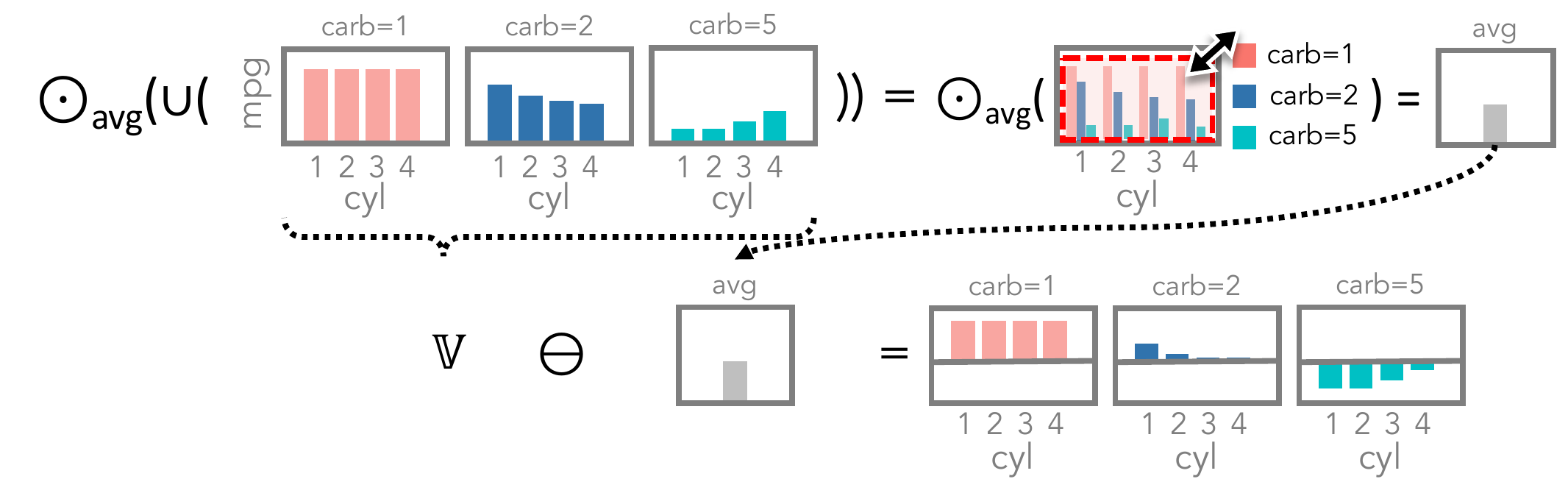}
  \vspace{-.2in}
  \caption{Alice computes the average delay across all carburetors, then subtracts the average from each carburetor chart.   }
  \label{f:iact_unionavg}
\end{figure}

\begin{example}
  Alice instead wants to compare each carburetor's fuel efficiency with the overall average across all carburetors and cylinders (\Cref{f:iact_unionavg}).
  She selects the three charts and juxtaposes their marks using $\cup$.
  She then selects all of the marks to compute their average using $\odot_{avg}$ (each mark is treated is an inidividual view).
  She finally drags the average onto the viewset to compute the differences.
  Note that the results differ from the previous example.
\end{example}

\begin{figure}[H]
  \centering
  \vspace{-1em}
  \includegraphics[width=\columnwidth]{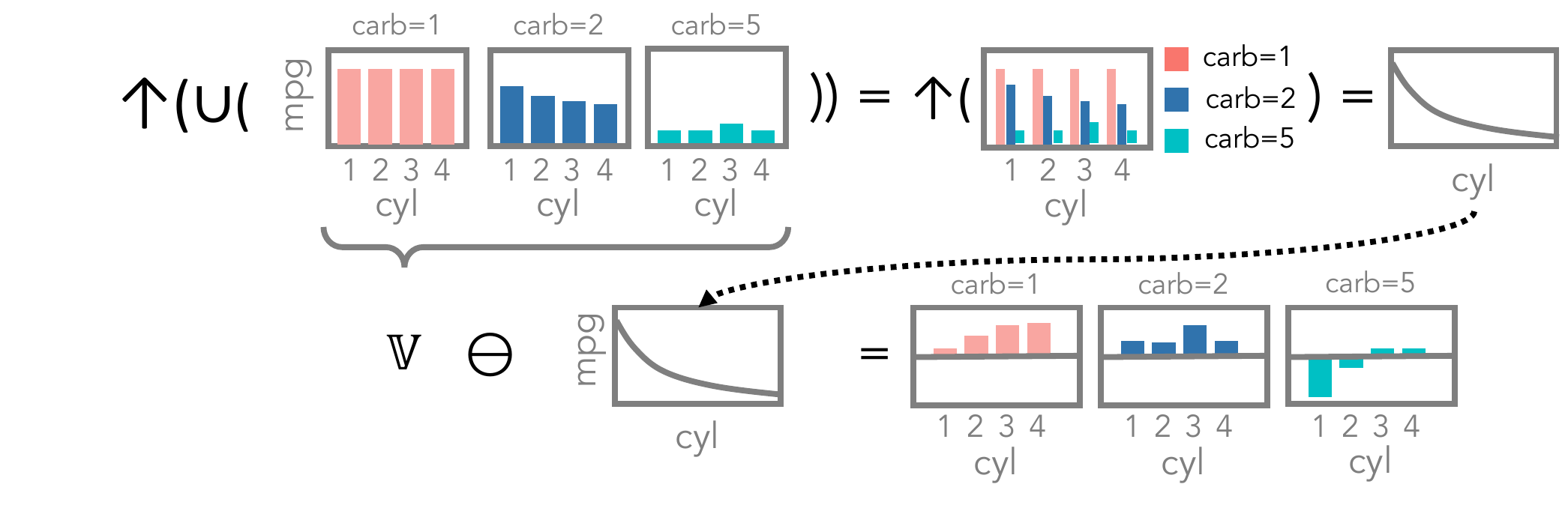}
  \vspace{-.2in}
  \caption{Alice fits a model to the mpg statistics across all carburetors, then plots the residuals for each carburetor plot.   }
  \label{f:iact_avgliftminus}
\end{figure}

\begin{example}
  Alice now wants to understand how each carburetor differs from the overal {\it expected} fuel efficiency
  for different cylinders (\Cref{f:iact_avgliftminus}).
  After she unions the three charts, she lifts the resulting view and doesn't
  select any of the labels in the legend.
  The model uses \texttt{cyl} for the features, and doesn't use a conditioning attribute.
  The resulting model view consists of a single model fitted using mpg statistics across all carburetors.
  Alice drags the model view to the viewset to complete the analysis.
\end{example}

In addition to its compositional properties, \sys
gives users flexibility to choose different comparison targets
based on their analysis needs.  
In the above examples, Alice could have just as easily have been analyzing other numbers of carburetors or other car attributes altogether, compared with the average delay across all US or European cars, or directly compared every pair of carburetor charts.

\subsection{\sys + View Manipulation and Coordination Interactions}

View manipulation and coordination interactions---such as sliders, pan and zoom, faceted navigation, linked brushing, coordinated views, and cross-filtering---update the data rendered in one of more views in response to user interactions. \sys naturally complements these interactions by enabling users to compare across interaction states.  

\begin{figure}
  \centering
  \includegraphics[width=\columnwidth]{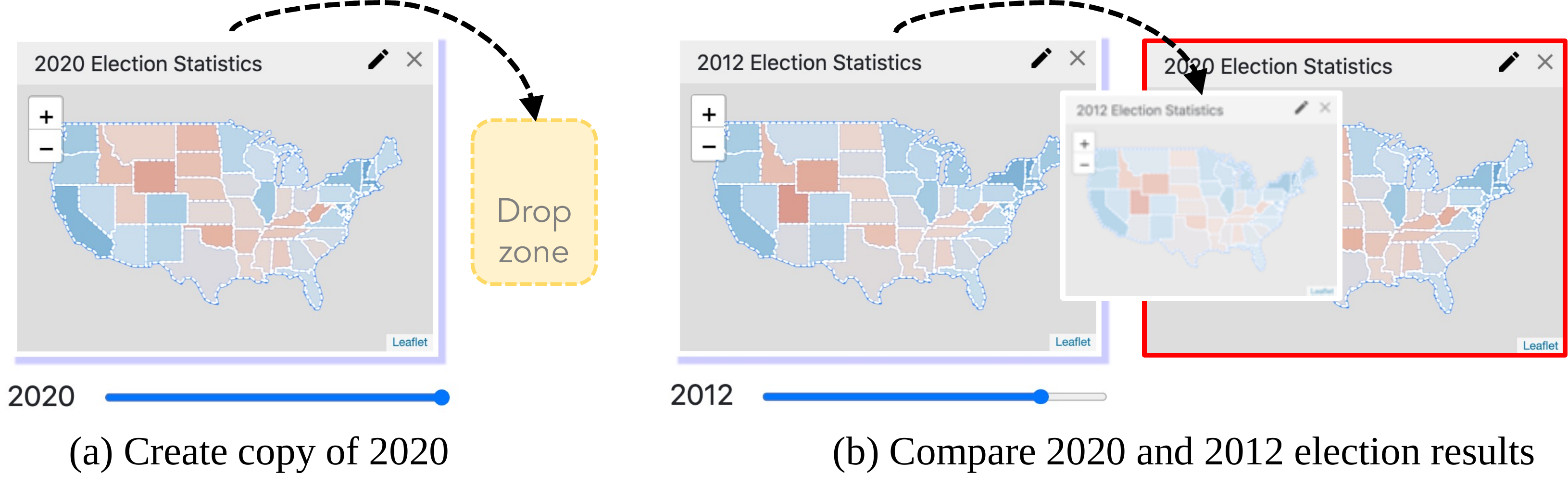}
  \vspace{-1em}
  \caption{Combining \sys with existing view interactions, such as a slider.  (a) The user drags the 2020 map to the drop zone, which creates a copy to the right of the slider.  (b) After she drags the slider to 2012, she can compare the 2012 map with the 2020 copy.}
  \label{f:exp_slider}
\end{figure}

\begin{example}
  \Cref{f:exp_slider} illustrates this principle using a slider.  The user drags the slider to show the 2020 election map, and drags the map to the ``Create View'' dropzone to create a copy. She then moves the slider to 2012, and compares that year's election map to the copy by dragging one on top of the other
\end{example}

This example does not use any additional functionality beyond view decomposition and composition described in this paper---it simply treats the slider's chart as a static view.   Thus, the approach is applicable to any interaction technique that changes the view's underlying data to a different group-by aggregation query.
  
A potential generalization of this technique builds upon visualization systems that automatically store provenance and/or interaction histories.  For instance, graphical histories~\cite{Heer2008GraphicalHF} track snapshots of past visualization state during data exploration,  optimistic visualizations~\cite{moritz2017trust} record approximate visualizations for users to later verify, and interaction snapshots~\cite{Wu2020FacilitatingEW} augment direct manipulation interfaces to record visualization requests.   All of these techniques automatically record and use past visualization state; \sys gives users the ability to compare against this historical state.

\subsection{Case Study Based on Google Finance}

Our final example presents a stack analysis application based on Google Finance,
which shows stock price information for a given company.  Although there are a multitude
of possible comparisons the user may want to make (e.g., how did yesterday's price compare with the past year?  
How do Google recent price compare with Amazon's average over last year?).
Our application shows how such an interface could be augmented to support comparison using companies and stock prices from the S\&P500.

\begin{example}
  \Cref{f:google_demo} shows details for Google and Amazon \circled{1},
    which includes its ticker, name, the current stock price, 
  and line chart of its recent per-day prices.
  Below the details is a list of cards that summarize related companies; each card shows the ticker and current stock price \circled{2}.
  All components that can be comparison targets have been highlighted in red, and the user can drag any target
onto any other target to compare them.  This displays the context menu \circled{3} where the user can choose, from left to right, to
compute their difference ($\ominus$), compute their average (treats the targets as a viewset and computes $\odot_{avg}$), 
or union them ($\cup$).  The comparison results are listed vertically on the right side.
For instance, \circled{4} shows the result of dragging Amazon's line chart onto Google's line chart and choosing ``Difference'';
\circled{5} is the result of dragging PIMCO's stock price onto Google's line chart and choosing ``Union''.
Finally, the comparison results are highlighted because they can also be used as comparison targets. 
\end{example}

\section{Discussion}\label{s:discussion}

Below, we discuss additional limitations and opportunities.

\stitle{Conceptual Model:} Any non-trivial visual analysis relies on non-trivial data analysis, and the primary observation in this paper is that comparison is a data- and design-level operation.  It relies on a clear separation between data transformation and visual encoding logic, where the data transformations are fully specified so that they can be fully analyzed.  This is in contrast with most existing visualization formalisms, which perform implicit data transformations that are not evident in the specification.  For instance, Vega-lite~\cite{Satyanarayan2017VegaLiteAG} supports transforms to pre-process the data before visual encoding, however encoding and scales specifications can implicitly perform binning and default aggregations.  This helps keep the specification concise, which simplifies the developer experience, but also makes comparison analysis more challenging.  Although Vega-lite compiles to the Vega data-flow graph~\cite{Satyanarayan2017VegaLiteAG}, {\it all logic}, including visual encoding specifications, are merged into a data flow representation that is too low-level to analyze.  An intermediate representation with a clean separation could enable broad application of comparison interactions.

\stitle{Relationship with Existing Interactions:}
Not all of the comparison operators are strictly necessary for a given visualization, and several in fact overlap with existing visualization interactions.  When the operations overlap, when are comparison interactions more effective?  We expect this to be context-dependent.  For instance, it is inefficient to create a standalone view to render a constant, only for the sake of subtracting it from a chart.  However, the same operation $\ominus$ helps to compare a scatterplot point (or table cell) with a chart.

These overlaps also help illuminate new properties for existing interactions and generalize their ideas.  For example, the union operator $\cup$ turns its input views into layers in the output view.  This is akin to layering charts in existing graphical grammars~\cite{Satyanarayan2017VegaLiteAG,wickham2016ggplot2,wilkinson2006grammar} and is already well-supported.  Although it may suggest that $\cup$ is redundant, it also suggests that layering is in fact a composition operation---taking safety into account can help avoid incorrect or confusing layering operations.  
In addition, the explode operator creates a small-multiples from a view and is akin to facetting, albiet with a subtle difference.  Explode facets a view that may be the result of a complex analysis and transformation process, and thus composes with other visual analysis operations.  In contrast, traditional facetting is typically used when creating a new view from the base data.

\stitle{Design:}
This paper proposes an interaction design to specify comparison targets and composition operators.  Its primary focus is to assess safety and derive the appropriate data transformations, and it uses simple rules to determine the output view's visual design.  We forsee three interesting directions to extend its design considerations.

First, is to extend the set of valid targets.  For instance, chart components such as axes or attribute orderings, encode dataset properies rather data directly, however they are clearly important for aiding comparison.  In addition, many visualizations are created from data flows more complex than the cube-operations described in this paper, and extending the formalism to e.g., D3 or pandas transformation programs helps extend comparison to a broader range of visual analysis settings.  

Second, is expand the language to support comparison tasks that require more sophisticated data and design strategies.  For instance, consider two time series line charts, costs by time and profits by time.  Although \sys can compute their difference over time or superimpose the trends together, another sensible comparison is to render a scatter plot of costs by profits.  However this would require complex data transformations---joining the two datasets by time and dropping the time attribute---and design transformations---choosing a point rather than line mark.  In addition, Gleicher~\cite{gleichercompare} also notes that the {\it scale} of a comparison necessitates both data strategies (filtering, summarization) and design strategies (scanning, interaction).  Of course, design considerations should only be incorporated into a comparison-specific language if they cannot be delegated to existing graphical grammars.  For instance, \sys supports side-by-side juxtaposition and not stacked charts because grammars like ggplot2~\cite{wickham2016ggplot2} can easily specify these designs using the \texttt{position} visual variable.

Third, is to better understand how comparison can be integrated into published visualizations and dashboards, which often have static layouts.  Our prototype creates a new view after every comparison, and simply renders all of them on the side.  How can visualizations easily accomodate new views, and when should comparisons update a view in place?   Or does comparison only make sense in the context of a multi-view exploration system?

\stitle{Relationship with Data Management: }
We described view composition as an instance of data integration~\cite{doan2012principles,batini1986comparative,Naumann2018SchemaM} that matches the attributes in the input views  (schema matching) and their values  (entity matching).  \sys uses conservative safety rules that only match identical attributes.  These rules could be extended by leveraging learned semantic attributes types~\cite{Zhang2020SatoCS} and adopting more flexible schema matching techniques~\cite{batini1986comparative,doan2012principles}.

Views render the results of a data processing workflow, and view composition is a meta-language over those results.  \sys is designed for SQL group-by aggregation workflows because they are widely used by visual analysis systems.  However, data science and scientific workflows both use more complex transformations, and process complex data types (e.g., 3D models) and relationships (networks, hierarchies).  Developing new composition semantics and the appropriate data integration techniques will be necessary for these workflows.

Finally, \sys compares and computes differences between data in the interface.  Another direction is to extend the formalism to support algorithms that explain~\cite{Wu2013ScorpionEA,Abuzaid2018DIFFAR,Roy2015ExplainingQA} the differences between comparison targets.

\stitle{Recommendation:}
An algebra offers several opportunities for composition recommendation. 1) Two views may not be safe to compose, but there may be intermediate data transformations or prior analysis steps that are safe to compose.  This is a way to search the analysis history for comparison candidates.  2) Similar to Scheidegger's visualizations by analogy~\cite{Scheidegger2007QueryingAC}, a user that wishes to compare an existing view with a newly imported dataset can be recommended the most likely sequence of transformations to arrive at a safe composition.   3) Existing view recommendation techniques~\cite{Lin2020DzibanBA,voyager,Kim2017GraphScapeAM,seedb} could be extended to recommend combinations of views to compose as the next analysis step.  4) Finally, multi-step composition can lead to a complex sequence of data transformations that is difficult to keep track of, and recommending informative view titles and descriptions can help users manage the views.

\section{Conclusion}

This paper presented View Composition Algebra (\sys), a formalism for composing entire of parts of visualizations to aid adhoc comparison tasks.  Users can select values, marks, legend elements, and entire charts as targets, and use composition operators to summarize or compare the targets.  We presented an interaction design, and used numerous examples to illustrate its novelty, efficacy, and ability to work alongside existing visual analysis interactions.   Finally, we argued that comparison goes beyond visual design, and also relies on data integration principles to assess comparison safety as well as complex data transformations to compute the comparisons.  Code is available at \url{http://viewcompositionalgebra.github.io}.

\acknowledgments{ The authors thank Remco Chang, Yunhai Wang, Michael Gleicher, and the reviewers for comments on earlier versions. This work was supported in part by NSF 1845638, 2008295, 2106197, 2103794.}

\balance
\bibliographystyle{abbrv-doi}
\bibliography{main}


\begin{table}[b]
    \begin{minipage}[b]{\linewidth}
      \parbox[b][][t]{.28\textwidth}{
      \includegraphics[width=.25\columnwidth]{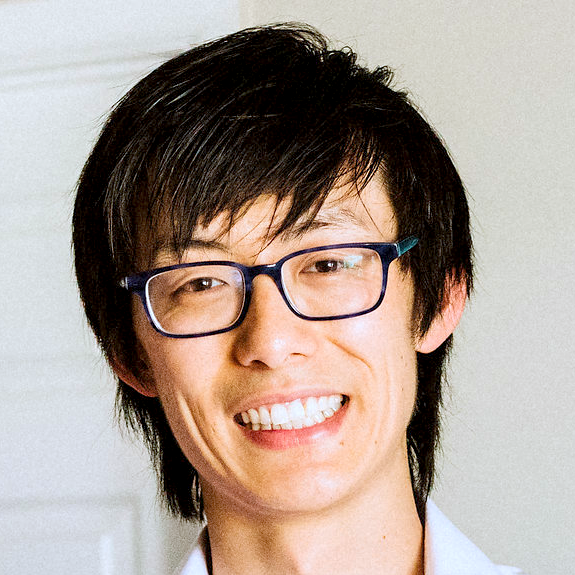}
    }
    \hfill
    \parbox[b][][t]{.71\textwidth}{ \textbf{Eugene Wu} is an Associate Professor at Columbia University. His research interests are in systems for human data management.  He has made contributions across the areas of database optimization, data visualization, data cleaning, stream processing systems, and crowd sourcing.  }
    \end{minipage}

\end{table}

\clearpage
\section{Appendix}

\subsection{View Composition Algebra}

\subsubsection{Statistical Composition with Nonexact Schemas}

We describe the safety rules for statistical composition $V^*=V_1\odot_{op} V_2$ in cases where the set of grouping attributes $A_{gb}^1$ in $Q_1$ 
is a strict super set of $A_{gb}^2$ in $Q_2$. 
\begin{align*}
  Q^* &= \pi_{A_{gb}^1, Q_1.y\ op\ Q_2.y\to y }(Q_1 \leftouterjoin_{A_{gb}^2} Q_2)\\
  R^* &= R_1
\end{align*}
$Q^*$ is defined over the {\it left} outer join of 
$Q_1$ and $Q_2$.  This join ensures that rows in $Q_1$ are preserved in the join
result, but rows in $Q_2$ that do not have a join match are not in the output.
We chose this because composition of views with different schemas is not 
symmetric---$V_1$ is transformed by matching data in $V_2$, but not vice versa.

How $A^2_{gb}$ is defined for $Q_2$ affects the semantics of the operator.  If $Q_2$ does not have any grouping attributes---it only contains a metric attribute---then every mark in $V_1$ will be composed with every metric value in $V_2$.  However, if $Q_2$ contains a \texttt{date} dimension, then each mark in $V_1$ will only be composed with the metric values in $V_2$ that have the same \texttt{date}.
Finally, note that the join condition is with respect to the smaller set of 
grouping attributes in $Q_2$,
whereas the attributes preserved in the query output are from $Q_1$.

\subsubsection{The Explode Operator}
The {\it explode} operator $\Xi_{A_e}(V)$ generates a set of views (a viewset), with one view for each group defined by the attributes $A_e$.   Explode is similar to facetting, which is traditionally used when creating new views from raw data.  In contrast, $\Xi$ is directly applied to existing views, including views derived from previous manipulations or compositions.  

The query for each output view first filters by the group, and then drops the explode attributes $A_e$; the visual mapping drops any mappings from $A_e$:
\begin{align*}
  \mathbb{Q}^* &= \{\pi_{\neg A_e}(\sigma_{A_e=vals}(Q)) | vals \in \gamma_{A_e}(Q)  \}\\
  R^* &= \{a\to a_v | a\to a_v\in R \land a\not\in A_e \}\\
  \mathbb{V}^* &= \{ R^*(Q^*) | Q^* \in \mathbb{Q}^* \}\\
      \neg A_e &= A_{gb} - A_e
\end{align*}

\subsubsection{ViewSet Union}
The union operator $V^*=\cup_{qid,a_t}(\mathbb{V})$ directly extends the 
binary union operator by extending and unioning all queries in the viewset,
and maps \hl{$qid$}  to an available visual attribute.
\begin{align*}
  Q'_i &= \pi_{Q_i.*, \hl{qid}}(Q_i(D))\hspace{2em}\forall i\in[1,n]\\
  Q^* & = Q'_1 \cup \ldots \cup Q'_n \\
  R^* &= \{qid\to\hl{$a$}\} \cup R_1\hspace{2em} s.t. \ a \in A_K - A_{R_1}
\end{align*}

\subsubsection{Viewset-View Composition}
Composition of viewsets and views returns a viewset, and corresponds to the cross product between the viewsets.  Below, let $\circ$ denote any binary operator:
\begin{align*}
  \mathbb{V} \circ V &= \{ V_i \circ V | V_i\in\mathbb{V} \}\\
  V \circ \mathbb{V} &= \{ V \circ V_i | V_i\in\mathbb{V} \}\\
  \mathbb{V}_1 \circ \mathbb{V}_2 &= \{ V_i \circ V_j | V_i\in\mathbb{V}_1, V_j\in\mathbb{V}_2 \}
\end{align*}

\subsection{Interaction Design}
\subsubsection{Operand Definitions}

We now formally define the operands for different chart components, using \Cref{f:iact_viewtypes} for reference.  Given the chart  $V=R(Q)$, we will define each view component as an operand $V^*$.  

\stitle{\circled{1} Entire View:} $V^* = V$.

\stitle{\circled{2} Labels in Legend:} 
Suppose the user selects the label $l$ for the grouping attribute $a_{sel}$, then the predicate is $a_{sel}=l$.   The resulting view drops $a_{sel}$ from the grouping attributes and visual mapping following the rules in \Cref{ss:algebra_stat}:
\begin{align*}
  Q^* &= \pi_{\lnot a_{sel}}\sigma_{a_{sel}=l}\left( Q \right) \\
  R^* &= \{ a_q\to a_v \in R | a_v \ne a_{sel}\} 
\end{align*}

\stitle{\circled{3} Marks:} $V^* = R(\sigma_p(Q))$, where predicate $p$ matches the selected marks.  For instance, the 1D selection in \Cref{f:iact_viewtypes}(a) would construct a predicate of the form $p: Date\in[min, max]$.

\stitle{Constant Value:}  $V^*=\pi_{val\to y}$, where $val$ is the constant.

\end{document}